\documentclass[journal]{IEEEtran}
\usepackage{graphicx}
\usepackage{amsfonts}
\usepackage{comment}
\usepackage{graphics} % for pdf, bitmapped graphics files
\usepackage{amsmath} % assumes amsmath package installed
\usepackage{amssymb}  % assumes amsmath package installed
\usepackage{enumitem}  %list item in the way of (i)(ii), see line 118
\usepackage{array}
\usepackage{nicematrix}
\usepackage{xcolor}
\usepackage{cite}

\usepackage{booktabs}
\usepackage{algorithm}
\usepackage{algorithmic}
\usepackage{balance}
\usepackage{caption}
\usepackage{subcaption}
\usepackage{amsthm}
\theoremstyle{definition}
\newtheorem{Definition}{Definition}
\newtheorem{Remark}{Remark}

\newtheorem{Lemma}{Lemma}
\newtheorem{Theorem}{Theorem}
\usepackage{nicematrix}
\usepackage[colorlinks,linkcolor=black,anchorcolor=black,citecolor=black,urlcolor=black]{hyperref}

\ifCLASSINFOpdf
\else
\fi

\begin{document}

\title{Optimal Honeypot Ratio and Convergent Fictitious-Play Learning in Signaling Games for CPS Defense}
%Cyber Deception Defense Based on Signaling Game and Fictitious Play
\author{Yueyue Xu, Yuewei~Chen, Lin Wang,~\IEEEmembership{Senior Member,~IEEE}, Zhaoyang Cheng, Xiaoming Hu,~\IEEEmembership{Senior Member,~IEEE}

%	\author{Yueyue Xu, Yuewei~Chen, Lin Wang, \textit{Senior Member, IEEE}, Zhaoyang Cheng, Xiaoming Hu, \textit{Senior Member, IEEE} % <-this % stops a space
	
	% \thanks{This paper is supported by ??? }	
    \thanks{Yueyue Xu is with the Department
of Automation, Shanghai Jiao
Tong University, Shanghai 200240, China
and also with KTH Royal Institute of Technology, Stockholm 10044, Sweden (e-mail: merryspread99@sjtu.edu.cn).}
	\thanks{Yuewei Chen and Lin Wang are with the Department
of Automation, Shanghai Jiao
Tong University, Shanghai 200240, China (e-mails: dave-c@sjtu.edu.cn, wanglin@sjtu.edu.cn). }% <-this % stops a space
\thanks{Zhaoyang Cheng and Xiaoming Hu are with the KTH Royal Institute of Technology, Stockholm 10044, Sweden (e-mails: zhcheng@kth.se, hu@kth.se).}
}

\markboth{IEEE Transactions on ...}
{Y. : Optimal Honeypot Ratio and Convergent Fictitious-Play Learning in Signaling Games for CPS Defense}

\maketitle

\begin{abstract}
Cyber–Physical Systems (CPSs) are facing a fast-growing wave of attacks.  
To achieve effective proactive defense, this paper models honeypot deployment as a $\gamma$-fixed signaling game in which node liveness serves as the only signal and normal-node signal $\gamma$ is exogenously fixed.  
We define the $\gamma$-perfect Bayesian–Nash equilibrium ($\gamma$-PBNE).  
Analytical expressions are obtained for all $\gamma$-PBNEs, revealing three distinct equilibrium regimes that depend on the priori honeypot ratio.  
Furthermore, the optimal honeypot ratio and signaling strategy that jointly maximize the network average utility are obtained.  
To capture strategic interaction over time, we develop a discrete-time fictitious-play algorithm that couples Bayesian belief updates with empirical best responses.  
We prove that, as long as the honeypot ratio is perturbed within a non-degenerate neighbourhood of the optimum, every fictitious-play path converges to the defender-optimal $\gamma$-PBNE.  
Numerical results confirm the effectiveness of the proposed method and demonstrate its applicability to CPS defense.
\end{abstract}


\IEEEpeerreviewmaketitle

%%%%%%%%%%%%%%%%%%%%%%%%%%%%%%%%%%%%%%%%%%%%%%%%%%%%%%%%%%%%%%%
%%%%%%%%%%%%%%%%%%%%%%%%%%%%%%%%%%%%%%%%%%%%%%%%%%%%%%%%%%%%%%%
% ---------- English Translation in LaTeX ----------
\section{Introduction}\label{sec1}

\IEEEPARstart{C}{yber}--Physical Systems (CPSs) are extensively deployed in critical
infrastructures such as transportation, power, healthcare, and
manufacturing.
In recent years, CPSs have suffered an explosive growth in
cyber-attacks~\cite{xing2024security}.
After surveying major industrial CPS incidents from 2000 to 2021,
Perera et~al.\ found that attacks typically adopt multi-stage
kill-chain tactics, and advanced threats can even span the entire
cyber-kill chain~\cite{kayan2022cybersecurity}.

Faced with a rapidly expanding attack surface, passive detection and patching alone are no
longer effective; proactive and deceptive defense is
attracting growing attention.
A honeypot is a deliberately exposed or emulated
server/host/service that appears indistinguishable from real assets
yet carries no critical workload.  Its purpose is to attract and log
intrusion activities so as to reveal attack strategies and patterns,
divert adversarial resources, and reduce the risk to genuine
systems~\cite{Spitzner2003Honeypots}.
Honeypots have evolved into two major families: low-interaction and high-interaction honeypots.Low-interaction honeypots are designed primarily for attack detection rather than in-depth analysis, which keeps their complexity and resource requirements low \cite{provos2004virtual}. In contrast, high-interaction honeypots emulate a full system environment to capture the complete attack chain and collect detailed attacker information \cite{garg2007deception}.

To model and analyze the interaction between the system defender and attacker, game theory offers an effective framework. In particular, signaling game has the merit of handling asymmetric and incomplete information, where the sender holds private information and conveys it through an observable signal, and then the receiver updates its beliefs via Bayes’ rule before acting. They therefore naturally fit network security scenarios in which attackers and defenders differ both in information and in timing. Typical applications include link-flooding attacks \cite{aydeger2021strategic}, co-resident attacks \cite{Hasan2020Signaling}, and moving-target defense strategies \cite{feng2017signaling}. 
Recent studies further embed probabilistic deception detectors into
signaling games, derive novel pooling and partially separating
equilibria, and quantify how detector performance affects strategic
outcomes~\cite{Pawlick2018Modeling}.
However, most existing models require the sender to explicitly
broadcast messages, which incurs additional communication cost; they also assume that the strategies of all nodes are
tunable, overlooking the fact that “normal” nodes in real networks are
constrained by operational duties and cannot change behaviour
arbitrarily.

Besides, most of the above signaling games research focus on static equilibria. However, the mere identification of equilibria does not guarantee that strategies of players will converge to them. Shifting to a dynamic perspective reveals a far more significant conclusions. Existing dynamical analyses for the signaling game include fictitious play, replicator dynamics, reinforcement learning and Moran processes\cite{huttegger2014some}. Among them, fictitious play provides a particularly appealing learning framework because it transplants the idea of best response directly into signaling games. Fudenberg and He \cite{fudenberg2018learning} embed the frequency-counting and the best-response logic of fictitious play into a large-population signaling environment: senders treat each signal as a multi-armed bandit, while receivers update empirical frequencies and best-respond accordingly. In the long run, the process selects only equilibria that satisfy the type-compatibility criterion. Building on this, Fudenberg et al. \cite{fudenberg2020payoff} allow agents to observe each other’s payoff functions, introducing the refinements of rationality-compatible equilibria (RCE) and unified RCE. They prove that Bayesian fictitious play converges almost surely to this refined equilibrium set.
Nevertheless, few studies integrate FP dynamics with Bayesian belief
updates to investigate strategic interaction and belief evolution
under incomplete information.

Motivated by these observations, we treat the liveness level of the node as the signal and assume that the liveness of the
normal nodes is exogenously fixed, thereby proposing a signaling
attack–defense game model that is closer to engineering reality.
Furthermore, we combine Bayesian updates with fictitious play to study the dynamics
of incomplete-information games.
Our core contributions are as follows:
\begin{enumerate}
    \item This article proposes a signaling game framework to model real-world cyber environments where the strategy of normal nodes $\gamma$ is fixed beforehand~\cite{pibil2012game}. Within this framework, $\gamma$-perfect Bayesian Nash equilibrium ($\gamma$-PBNE) is introduced and derived.

    \item The optimal defense strategy based on the optimal $\gamma$-PBNE is established. By defining the network average utility, we determine both the optimal honeypot ratio and the optimal equilibrium strategy that jointly maximize the network average utility.

    \item A discrete-time fictitious-play learning algorithm for the $\gamma$-fixed signaling game is developed, which studies the strategy dynamics under asymmetric information. It is proved that when the defender perturbs the honeypot ratio within a non-degenerate neighbourhood around the optimal value, the fictitious-play trajectory converges to the optimal $\gamma$-PBNE.
\end{enumerate}

The remainder of this paper is organized as follows:
Section~\ref{sec2} introduces the signaling game model;
Section~\ref{sec3} derives the perfect Bayesian Nash equilibria; Section~\ref{sec4} gives the optimal defense strategy; Section~\ref{sec5} establishes a fictitious-play learning algorithm and gives the convergence condition;
Section~\ref{sec6} validates the theoretical results through
simulation; finally, Section~\ref{sec7} concludes this paper.

%%%%%%%%%%%%%%%%%%%%%%%
%%%%%%%%%%%%%%%%%%%%%%%%%%%%%%%%%%%%%%%%%%%%%%%%%%%%%%%%%%%%%%%
\begin{table}[h]
\centering
\caption{Summary of Notation}\label{tab:notation}
\begin{tabular}{@{} >{\centering\arraybackslash}m{2.4cm}  c @{}}
    \toprule
    Notation & Meaning \\
    \midrule
    $D, A$ & Defender, Attacker\\
            $\theta \! \in \Theta,\!m \in \! \mathbb{M}, a \! \in \! \mathbb{A}$ & Types, Messages, Actions \\
    $u_i(\theta, m, a)$ & Utility Functions of Player $i \in \{ D,A\}$ \\
    $\sigma_D(m \mid \theta)$ & Signaling Strategy of $D$ of Type $\theta$ \\
    $\sigma_A(a \mid m)$ & Attack Strategy of $A$ given $m$ \\
    $p$ & Prior Probability of Type $\theta_1$ \\
    $\mu_A(\theta \mid m)$ & Posterior Belief of $A$ that $D$ is of Type $\theta$ \\
    $\overline{U}_{net}$ & Network Average Utility Function\\
    $M^*$ & Optimal Number of Honeypots\\
    $\alpha$    &   Payoff of a honeypot sending L when attacked   \\
    $f\alpha$   &   Payoff of a honeypot sending H when attacked   \\
    $-g\alpha$  &   Payoff of a normal node sending L when attacked\\
    $-hg\alpha$ &   Payoff of a normal node sending H when attacked \\
    $\beta$     &   Honeypot maintenance cost      \\
    $c_d$       &   Cost of sending H by honeypot \\
    $c_a$       &   Attack cost      \\
    \midrule
    \end{tabular}
\end{table}
\section{Signaling Game Model}\label{sec2}
A signaling game $\mathcal{G}^0$ is introduced to model the network attack and defense scenario. Each node is considered as a defender (D) which acts as the signal sender, while the signal receiver which acts as the attacker (A) can choose whether to attack each node or not. 

%%%%%%%%%%%%%%%%%%%%%%%%%%%%%%%%%%%%%%%%%%%%%%%%%%%%%%%%%%%%%%%
\subsection{Types, Messages, Actions, and Beliefs}

Table \ref{tab:notation} summarizes the notation. Firstly we define the type of the defender as $\theta \in \Theta=\{\theta_1, \theta_2\}$, where type $\theta_1$ is a honeypot and type $\theta_2$ is a normal node. The type is drown from a probability distribution, i.e.,
\begin{equation*}
\Pr(\theta_1)=p, \Pr(\theta_2)=1-p,
\end{equation*}
where $\Pr()$ is the probability function.

Based on the type, defender chooses messages $m \in \mathbb{M} = \{H, L\}$, representing high liveness and low liveness respectively.
Define the strategy of defender as
\begin{equation}
\sigma_D
=\begin{bmatrix}
\sigma_D(H\mid \theta_1) & \sigma_D(L\mid \theta_1)\\[4pt]
\sigma_D(H\mid \theta_2) & \sigma_D(L\mid \theta_2)
\end{bmatrix} \in \Gamma_D,
\end{equation}
where $\sigma_D(m \mid \theta) \in \mathbb{R}$ gives the probability with which the defender sends message $m$ given that it is of type $\theta$, $\Gamma_D  \in \mathbb{R}^{2\times 2}$ is the space of strategies defined as $\Gamma_D = \left\{ \sigma_D \mid \forall \theta, \sum_{m \in \mathbb{M}} \sigma_D(m \mid \theta) = 1;
\ \forall \theta, m, \ \sigma_D(m \mid \theta) \geq 0 \right\}$.

Next, the attacker receives message $m$, and chooses an action $a \in \mathbb{A}=\{A, N\}$, representing attack and not attack. Define the strategy of the attacker as 
\begin{equation}
\sigma_A=\begin{bmatrix}
\sigma_A(A\mid H) & \sigma_A(N\mid H)\\[4pt]
\sigma_A(A\mid L) & \sigma_A(N\mid L)
\end{bmatrix}  \in \Gamma_A,
\end{equation}
where $\sigma_A(a \mid m) \in \mathbb{R}$ is the probability of playing action $a$ given message $m$, $\Gamma_A \in \mathbb{R}^{2\times 2}$ is the space of strategies defined as 
$\Gamma_A =\left\{ \sigma_A \mid \forall m,  \sum_{a \in \mathbb{A}} \sigma_A(a \mid m) = 1; \ \forall  m, a, \ \sigma_A(a \mid m) \geq 0 \right\}$.

Based on the received signal $m$, the attacker forms a belief $\mu_A(\theta|m),\, \theta \in \Theta$ about the type $\theta$ of defender, where $\mu_A(\theta|m)$ is the probability that the attacker believes the defender is of type $\theta$ and $\sum_{\theta \in \Theta} \mu_A(\theta|m) = 1$. 
The attacker uses posterior belief $\mu_A(\theta|m)$ to decide actions.

\begin{figure*}[!htp]
    \centering
    \includegraphics[width=0.8\textwidth]{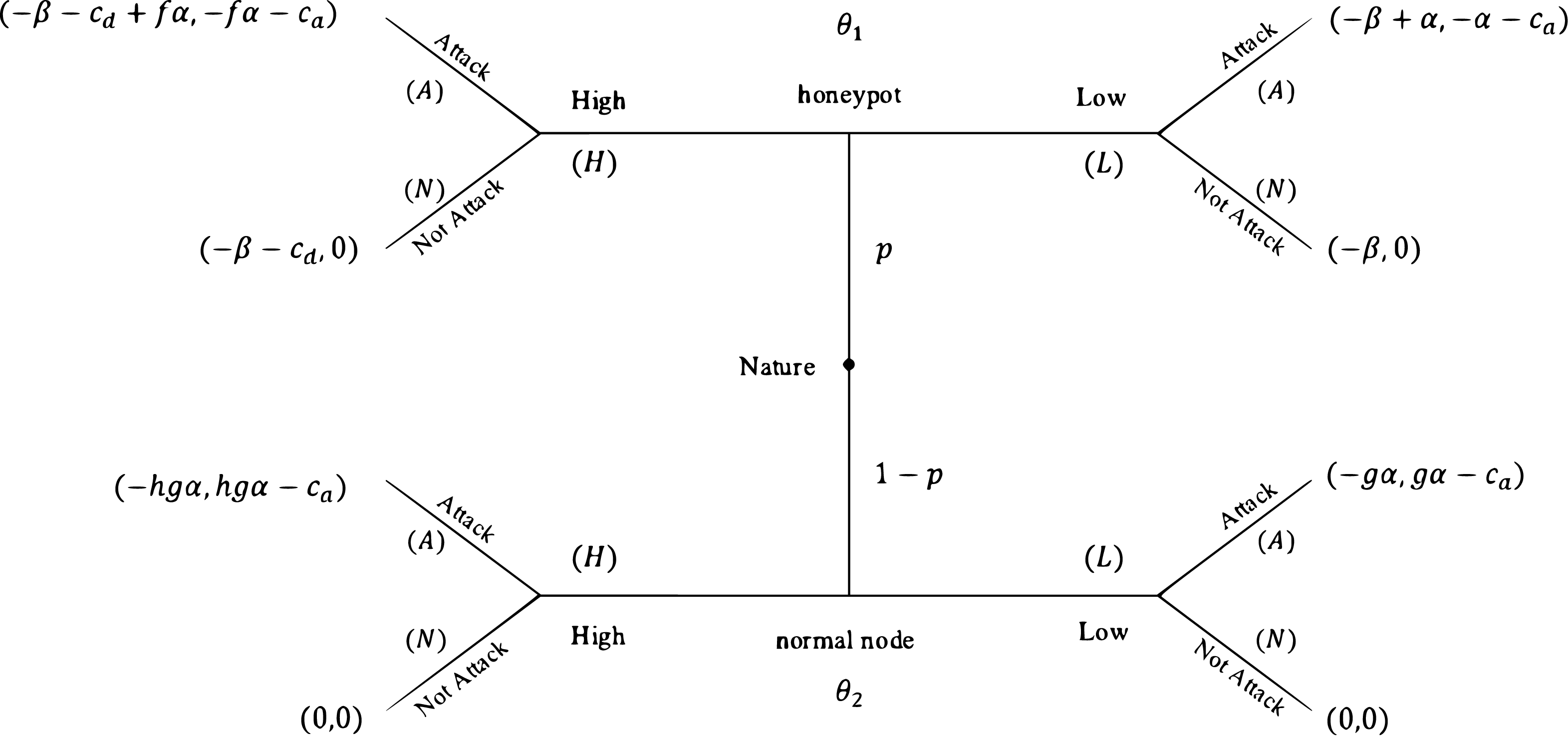}
    \caption{Signaling game model representation $\mathcal{G}^0$, with \((u_D,u_A)\) pairs displayed at the terminal nodes.}
    \label{fig:model}
\end{figure*}
%%%%%%%%%%%%%%%%%%%%%%%%%%%%%%%%%%%%%%%%%%%%%%%%%%%%%%%%%%%%%%%
\subsection{Utility Functions}
Let 
$u_i:\Theta\times \mathbb{M}\times\mathbb{A}\to\mathbb{R}$, for $i\in\{D,A\}$, denote the utility functions for the defender ($i = D$) or the attacker ($i = A$). Consequently,
$u_i\bigl(\theta, m, a\bigr)$
gives the payoff to player $i$ when the type of the defender is $\theta \in\Theta$, the defender sends message $m\in\mathbb{M}$, and the action of the attacker is $a\in\mathbb{A}\,. $

We first define some parameters involved in the utility functions, which is summarized in Table~\ref{tab:notation}. All parameters are positive. $\alpha$, $f\alpha$, $-g\alpha$, $-hg\alpha$ represent the returns of the defender under attack when it is a low-liveness honeypot, a high-liveness honeypot, a low-liveness normal node, and a high-liveness normal node, respectively. In each case, the return of the attacker is simply the negative of the return of the defender. 
We find that when a honeypot is attacked, the defender gains a positive return, whereas the attacker incurs a negative return.
This is because, when a honeypot is under attack, the system itself remains unharmed and can collect valuable information about the attacker. In contrast,
when a normal node is attacked, the defender gains a negative return, whereas the attacker receives a positive return.
This is due to the damage the system incurs, which varies according to the node’s liveness.
Furthermore, \(\beta\) is the cost incurred by the defender 
for maintaining a honeypot, \(c_d\) is the additional cost for operating a high-liveness honeypot, 
and \(c_a\) is the cost borne by the attacker when launching an attack. 

Additionally, to ensure the reasonableness of the game model, we establish several constraints on the parameters, which are summarized in Table~\ref{tab:utility_constraints}.
\begin{table}[!hpt]
  \caption{Assumptions for utility function parameters}
  \label{tab:utility_constraints}
  \centering
  \begin{tabular}{cl}
    \toprule
    Assumption    &   Meaning  \\
    \midrule
    $f>1$    &   A high-liveness honeypot gets more \\
             & attacker information than a low-liveness one.\\ \hline
    $h>1$    &   A high-liveness normal node suffers greater \\
             &   losses when attacked than a low-liveness one. \\ \hline
    $-\beta + \alpha > 0$  
             &   A low-liveness honeypot gains net benefit \\
             &   from being attacked. \\ \hline
    $-\beta - \!c_d \!+ \!f \alpha \! > \!-\beta\! + \!\alpha$  
             &   A high-liveness honeypot has higher utility \\
             &  under attack than a low-liveness one. \\ \hline
    $g\alpha - c_a > 0$    
             &   Attacking a normal node yields net benefit \\
             &   for the attacker. \\ 
    \bottomrule
  \end{tabular}
\end{table}

Figure~\ref{fig:model} illustrates the signaling game for the cyber attack and deception defense. 
The figure shows all eight possible outcomes (honeypot or normal node, high or low signal, attack or not attack) and the resulting \((u_D,u_A)\) payoffs at the terminal nodes.
For example, consider the case in which the defender is a honeypot (\(\theta_1\)), sends a high signal (\(H\)), and the attacker chooses to attack (\(A\)). This corresponds to the top-left node in Figure~\ref{fig:model}. In this case, because the high-liveness honeypot is attacked, the defender gains a benefit $f\alpha$, while the attacker suffers a loss of $f\alpha$. 
Furthermore, the defender has the cost $\beta$ for maintaining the honeypot and the extra cost $c_d$ for maintaining high-liveness. Thus, the total utility for the defender is $u_D(\theta_1, H, A)=-\beta - c_d+f \alpha$. 
The attacker has the cost $c_a$ for choosing to attack and the total utility for the attacker is $u_A(\theta_1, H, A)=-f\alpha - c_a$. Utilities of the defender and attacker for the other cases are defined in a similar manner.

Define an expected utility function $U_D: \Gamma_D \times \Gamma_A \to \mathbb{R}$ such that $U_D(\sigma_D,\sigma_A\mid \theta)$ gives the expected utility to the defender when it plays strategy $\sigma_D$, given that she is of type $\theta$. This expected utility is given by
\begin{equation}
U_D(\sigma_D,\sigma_A\mid \theta)
=\sum_{a\in \mathbb{A}}\sum_{m\in \mathbb{M}}
\sigma_A(a\mid m)\,\sigma_D(m\mid \theta)\,u_D(\theta,m,a).
\end{equation}
Next define $U_A: \Gamma_A \to \mathbb{R}$ such that $U_A(\sigma_A\mid \theta,m)$ gives the expected utility to the attacker when he plays strategy $\sigma_A$ given message $m$ and sender type $\theta$. The expected utility function is given by
\begin{equation} 
U_A(\sigma_A\mid \theta,m) =\sum_{a\in \mathbb{A}}\sigma_A(a\mid m)u_A(\theta,m,a). 
\end{equation}
%%%%%%%%%%%%%%%%%%%%%%%%%%%%%%%%%%%%%%%%%%%%%%%%%%%%%%%%%%%%%%%
\subsection{Equilibrium Concept}

Perfect Bayesian Nash equilibrium (PBNE) is often used to analyze the equilibrium cases of signaling games. In most research on signaling games, researchers concentrate on pure-strategy PBNEs, which are solved by enumerating candidate strategies and checking whether each is consistent with the beliefs along the equilibrium path \cite{aydeger2021strategic}. Because pure-strategy equilibria can be nonexistent or excessively rare, we also examine mixed-strategy equilibria in this study. 

Here follows the definition of PBNE for both pure and mixed strategies. A PBNE is a sender strategy \(\sigma_{D}^{*}\), a receiver strategy \(\sigma_{A}^{*}\) and beliefs \(\mu_{A}(\theta \mid m)\) such that: 

(i) given \(\sigma_{A}^{*}\), for every type $\theta$, \(\sigma_{D}^{*}(m \mid \theta)\) maximizes the expected utility of the defender $U_D(\sigma_D,\sigma_A^{*}\mid \theta)$; 

(ii) given \(\mu_{A}(\theta \mid m)\), for every message \(m\), \(\sigma_{A}^{*}(a \mid m)\) maximizes the expected utility of the attacker $\sum_{\theta \in \Theta}\mu_A(\theta\mid m)\,U_A\bigl(\sigma_A\mid \theta,m\bigr)$; 

(iii) \(\mu_{A}(\theta\mid m)\) is derived from Bayes’ rule for any message sent with positive probability and is otherwise arbitrary so long as the strategy of the receiver remains optimal. 

When both strategies \(\sigma_{D}^{*}\), \(\sigma_{A}^{*}\) place probability one on single actions, the equilibrium is pure strategy PBNE \cite{gibbons1992primer}; otherwise it is mixed strategy PBNE \cite{fudenberg1991game}.

However, in real‑world cyber environments, the signal level transmitted by a normal node $(\theta_2)$ is not configurable but is determined by the network’s actual operational conditions\cite{pibil2012game}. 
Following the concept of a commitment type in the reputation game (\cite{fudenberg1992maintaining, pei2022reputation}), we assume that the normal node is a commitment type—its signaling strategy is exogenously fixed. 
%The key distinction in our model is that this commitment type participates in a one‐shot signaling game rather than in the infinitely repeated interactions of the canonical reputation framework.
Specifically, we fix $\sigma_{D}(H\mid \theta_2)=\gamma \in [0,1]$, so that a normal node sends the high signal \(H\) with probability \(\gamma\) and the low signal \(L\) with probability \(1-\gamma\).

Consequently, we need to redefine PBNE in this case, which we call \(\gamma\)-PBNE. Unlike in a standard PBNE, the strategy of the type-\(\theta_{2}\) defender is not optimized in a \(\gamma\)-PBNE.
We now define two equilibria: the \(\gamma\)-Pure PBNE, where all players except the normal node use pure strategies, and the \(\gamma\)-Mixed PBNE, where at least one such player mixes his strategy.
\begin{Definition}
\label{def:gamma-pure-PBNE}
Assume the strategy of the normal nodes is fixed at 
$\sigma_{D}(H\mid \theta_2)=\gamma$, where $\gamma\in(0,1)$. 
A $\gamma$-Pure PBNE of the signaling game is a strategy profile $(m^*(\theta_1),a^*(m))$ and posterior beliefs $\mu_A(\theta\mid m)$ such that
\begin{equation}
m^*(\theta_1)\in\arg\max_{m \in \mathbb{M}}u_D\bigl(\theta_1,m,a^*(m)\bigr),
\end{equation}
\begin{equation}
\forall m \in \mathbb{M},\quad
a^*(m)\in\arg\max_{a \in \mathbb{A}}\sum_{\theta \in \Theta}\mu_A(\theta\mid m)\,u_A(\theta,m,a),
\end{equation}
with $\sum_{\theta \in \Theta}\mu_A(\theta\mid m)=1$, where
\begin{equation}\label{pure_belief}
\mu_A(\theta\mid m_j)
=\frac{\Pr(\theta)}
      {\sum_{\tilde \theta \in\Theta_j}\,\Pr(\tilde \theta)},
\end{equation}
where $\Theta_j$ denotes the set of types that send the message $m_j$.
\end{Definition}

\begin{Definition}
\label{def:gamma-mixed-PBNE}
Assume the strategy of the normal nodes is fixed at 
$\sigma_{D}(H\mid \theta_2)=\gamma$, where $\gamma\in[0,1]$. 
A $\gamma$-Mixed PBNE of the signaling game is a profile $(\sigma_D^*(m\mid \theta_1),\sigma_A^*)$ and posterior beliefs $\mu_A(\theta\mid m)$ such that
\begin{equation}
\sigma_D^*(m\mid \theta_1)\in\arg\max U_D\bigl(\sigma_D(m\mid \theta_1),\sigma_A^*\mid \theta_1\bigr), 
\end{equation}
\begin{equation}
\forall m\in\mathbb{M},\quad
\sigma_A^*\in\arg\max_{\sigma_A\in\Gamma_A}\sum_{\theta \in\Theta}\mu_A(\theta\mid m)\,U_A\bigl(\sigma_A\mid \theta,m\bigr),
\end{equation}
with $\sum_{\theta \in\Theta}\mu_A(\theta\mid m)=1$.  If 
$\displaystyle\sum_{\tilde \theta \in\Theta}\sigma_D(m\mid\tilde \theta)\,\Pr(\tilde \theta)>0$, then
\begin{equation}\label{mixed_belief}
\mu_A(\theta\mid m)
=\frac{\sigma_D(m\mid \theta)\,\Pr(\theta)}
      {\sum_{\tilde \theta \in\Theta}\sigma_D(m\mid\tilde \theta)\,\Pr(\tilde \theta)},
\end{equation}
otherwise $\mu_A(\theta \mid m)$ may be any probability distribution over~$\Theta$.
\end{Definition}
\begin{Remark}
In this paper, we do not classify equilibria according to their information‑disclosure patterns (separating, pooling, or partially‑separating). This is because when the strategy of type‑$\theta_2$ defender is fixed as mixed strategy $\gamma$, $\gamma\in(0,1)$, only partially‐separating equilibria can exist. Furthermore, we consider mixed‐strategy equilibria because pure‐strategy equilibria are relatively simplistic and may fail to achieve the defense effect we seek, which we will discuss below.
\end{Remark}
%%%%%%%%%%%%%%%%%%%%%%%%%%%%%%%%%%%%%%%%%%%%%%%%%%%%%%%%%%%%%%%
%%%%%%%%%%%%%%%%%%%%%%%%%%%%%%%%%%%%%%%%%%%%%%%%%%%%%%%%%%%%%%%
\section{Equilibrium Analysis}\label{sec3}
In this section, we will analyze PBNEs when the strategy of the normal nodes is fixed. 
Since the posterior beliefs $\mu_A(\theta|m) $ according to \eqref{pure_belief} and \eqref{mixed_belief} include two elements--$\mu_A(\theta_1|H) $ and $\mu_A(\theta_1|L)$--represent the belief for the type of the defender when receiving signals $H$ and $L$, respectively. 
To simplify notations, we define 
\begin{equation}\label{mu_HL}
\mu_A(\theta_1|H)=\mu_H,\quad \mu_A(\theta_1|L)=\mu_L.
\end{equation}
The equilibrium beliefs presented below will be represented using $\mu_H$ and $\mu_L$.
%%%%%%%%%%%%%%%%%%%%%%%%%%%%%%%%%%%%%%%%%%%%%%%%%%%%%%%%%%%%%%%
%%%%%%%%%%%%%%%%%%%%%%%%%%%%%%%%%%%%%%%%%%%%%%%%%%%%%%%%%%%%%%%
\subsection{Pure strategy PBNE}
In this subsection, we analyze the $\gamma$‑Pure PBNE of the signaling game $\mathcal{G}^0$, which is defined in Definition \ref{def:gamma-pure-PBNE}. The next theorem characterizes the $\gamma$-Pure PBNE.
\begin{Theorem}\label{thm2}
In the signaling game $\mathcal{G}^0$, given that the strategy of the normal node ($\theta_2$) is fixed as $\sigma_{D}(H\mid \theta_2)=\gamma,~\gamma\in (0,1)$, there exists a unique $\gamma$‑Pure PBNE when $p \leq p_1= \frac{\gamma(hg\alpha-c_a)}{\gamma(hg\alpha-c_a) +f\alpha +c_a}$ as below:
\begin{equation}\label{pbne:Hd_2AA}
\left\{ \sigma_D \!=\! \begin{bmatrix}
    1 & 0 \\ \gamma & 1-\gamma
\end{bmatrix}, \sigma_A\! =\! \begin{bmatrix}
    1 & 0\\
    1 & 0
\end{bmatrix}, \mu_H\! = \!\frac{p}{p + (1-p) \gamma}, \mu_L \!=\! 0 \right\},
\end{equation}
where $\mu_H$ and $\mu_L$ are equilibrium beliefs defined in \eqref{mu_HL}.
\end{Theorem}
\begin{IEEEproof}
See Appendix A.
\end{IEEEproof}

Theorem~\ref{thm2} indicates that when $p \leq p_1$, there exists an equilibrium where the honeypot always sends $H$, and the attacker always chooses $A$ regardless of the signal type.
For the scenario where $p \geq p_1$, the likelihood of the defender being a honeypot ($\theta_1$) is higher. Since choosing action $A$ against a honeypot reduces the utility of the attacker, the attacker has an incentive to deviate from $A$, and thereby the equilibrium \eqref{pbne:Hd_2AA} is disrupted.

%%%%%%%%%%%%%%%%%%%%%%%%%%%%%%%%%%%%%%%%%%%%%%%%%%%%%%%%%%%%%%%
\subsection{Mixed strategy PBNE}
By Theorem \ref{thm2}, when $p> p_1$, no $\gamma$‑Pure PBNE exists. 
Several studies on signaling games analyze mixed-strategy equilibria when no pure-strategy equilibrium exists (e.g., \cite{huttegger2014some, ropero2025signaling}). The standard procedure for pure-strategy equilibria is to posit separating, pooling, or semi-separating outcomes and then verify the corresponding incentive constraints, as shown in the Appendix A. This approach, however, does not apply directly to mixed-strategy equilibria.
Accordingly, in this subsection, we further analyze the $\gamma$-Mixed PBNE, which is defined in Definition \ref{def:gamma-mixed-PBNE}. The next theorem characterizes the $\gamma$-Mixed PBNE.

\begin{Theorem}\label{thm:mixed_PBNE}
In the signaling game $\mathcal{G}^0$, given that the strategy of the normal node ($\theta_2$) is $\sigma_{D}(H\mid \theta_2) = \gamma,~\gamma\in (0,1)$, there exists different mixed strategy equilibrium with different type probability $p \in (0,1)$:
\\(i) When $0<p < p_1$, the mixed strategy equilibrium is
\begin{equation}\label{mix_eq1}
\left\{\! \sigma_D \!=\! \begin{bmatrix}
    1 & 0 \\ \gamma & 1-\gamma
\end{bmatrix}, \sigma_A\! =\! \begin{bmatrix}
    1 & 0\\
    1 & 0
\end{bmatrix}, \mu_H\! = \!\frac{p}{p + (1-p) \gamma}, \mu_L \!=\! 0 \!\right\};
\end{equation}
\\(ii) When $p_1 <  p < p_2$, the mixed strategy equilibrium is
\begin{multline}\label{mix_eq2}
\left\{ \sigma_D \!=\! \begin{bmatrix}
    F_1 & 1-F_1 \\ \gamma & 1-\gamma
\end{bmatrix}, \sigma_A\! =\! \begin{bmatrix}
    \frac{\alpha+c_d}{f\alpha} & 1-\frac{\alpha+c_d}{f\alpha}\\
    1 & 0
\end{bmatrix},
\right.\\
    \left. \mu_H\! = \!\frac{p F_1}{p F_1+ (1-p) \gamma}, \mu_L\!=\!\frac{p (1-F_1)}{p (1-F_1)+ (1-p) (1-\gamma)}\right\};
\end{multline}
\\(iii) When $p_2 < p<1$, the mixed strategy equilibrium is
\begin{multline}\label{mix_eq3}
\left\{ \sigma_D \!=\! \begin{bmatrix}
    F_1 & 1-F_1 \\ \gamma & 1-\gamma
\end{bmatrix}, \sigma_A\! =\! \begin{bmatrix}
    \frac{c_d}{f\alpha} & 1-\frac{c_d}{f\alpha}\\
    0 & 1
\end{bmatrix},
\right.\\
    \left. \mu_H\! = \!\frac{p F_1}{p F_1+ (1-p) \gamma}, \mu_L \!=\!\frac{p (1-F_1)}{p (1-F_1)+ (1-p) (1-\gamma)}\right\};
\end{multline}
\\(iv) When $p=p_1$, the mixed strategy equilibrium is
\begin{multline}\label{mix_eq4}
\left\{ \sigma_D \!=\! \begin{bmatrix}
    1 & 0 \\ \gamma & 1-\gamma
\end{bmatrix}, \sigma_A\! =\! \begin{bmatrix}
    a_H^* & 1-a_H^*\\
    1 & 0
\end{bmatrix},
\right.\\
    \left.  a_H^* \in [\frac{\alpha + c_d}{f \alpha},1], \mu_H\! = \!\frac{p}{p + (1-p) \gamma},\mu_L \!=\!0\right\};
\end{multline}
\\(v) When $p=p_2$, the mixed strategy equilibrium is
{\scriptsize
\begin{multline}\label{mix_eq5}
\left\{ \sigma_D \!=\! \begin{bmatrix}
    F_1 & 1-F_1 \\ \gamma & 1-\gamma
\end{bmatrix}, \sigma_A\! =\! \begin{bmatrix}
    a_H^* & 1-a_H^*\\
    (a_H^* f - \frac{c_ d}{\alpha}) & (1-a_H^* f + \frac{c_ d}{\alpha})
\end{bmatrix}, 
\right.\\
    a_H^*\! \in\! [\frac{c_d}{f \alpha},\! \frac{\alpha + c_d}{f \alpha}],\left.\!  \mu_H\! = \!\frac{p F_1}{p F_1+ (1-p) \gamma}, \mu_L\!=\!\frac{p (1-F_1)}{p (1-F_1)+ (1-p) (1-\gamma)}\! \right\};
\end{multline}}
where 
\begin{equation}\label{F_1}
F_1= \frac{hg\alpha -c_a}{f\alpha + c_a} \cdot \frac{1-p}{p} \cdot \gamma, 
\end{equation}
\begin{equation}\label{theta12}
p_{1}=\frac{A \gamma}{1+A \gamma},\, p_{2}=\frac{A \gamma+B(1-\gamma)}{1+A \gamma+B(1-\gamma)}, 
\end{equation}
\begin{equation}\label{AB}
A=\frac{h g \alpha-c_a}{f \alpha+c_a}>0, \quad B=\frac{g \alpha-c_a}{\alpha+c_a}>0 .
\end{equation}
\end{Theorem}
\begin{IEEEproof}
We parameterize the strategies of the defender and attacker as follows
\begin{equation}
\sigma_D = \begin{bmatrix} d_1 & 1-d_1 \\ \gamma & 1-\gamma \end{bmatrix},~ \sigma_A = \begin{bmatrix}a_H & 1-a_H\\ a_L & 1- a_L\end{bmatrix},
\end{equation}
where $d_1,a_H, a_L \in [0,1]$. Based on equations~\eqref{mixed_belief} and \eqref{mu_HL}, we can easily calculate the posterior belief $\mu_H$ and $\mu_L$:
\begin{align}
\mu_H & = \frac{p d_1}{p d_1 + (1-p) \gamma},\label{PBNE_p}\\
\mu_L & = \frac{p (1-d_1)}{p (1-d_1) + (1-p) (1-\gamma)}.\label{PBNE_q}
\end{align}

First of all, we calculate the expected utility of the attacker $U_A(\theta,H,a)$ when receiving signal $H$:
\begin{align}
&U_A(\theta,H,a)  = \sum_{\theta \in \Theta} \sum_{a \in \mathbb{A}} \mu_A(\theta|H) \cdot \sigma_A(a|H) \cdot u_A(\theta,H,a) \nonumber\\
& =a_H \cdot \mu_A(\theta_1|H) \cdot (-f\alpha - c_a) +a_H \cdot \mu_A(\theta_2|H) \cdot (hg\alpha -c_a) \nonumber\\
& =a_H \cdot \frac{p d_1(-f\alpha - c_a) + (1-p) \gamma (hg\alpha -c_a)}{p d_1 + (1-p) \gamma}.
\end{align}
Maximizing $U_A(\theta,H,a)$ by $a_H$, we obtain:
\begin{subequations}\label{eq:r*}
\begin{align}
a_H^* = 1, &\,\text{when } d_1< F_1,\label{r*_a}\\
a_H^* = 0, &\,\text{when } d_1> F_1,\label{r*_b}\\
a_H^* \in  [0,1], &\,\text{when } d_1= F_1,\label{r*_c}
\end{align}
\end{subequations}
where
\begin{align}
F_1  = \frac{hg\alpha -c_a}{f\alpha + c_a} \cdot \frac{1-p}{p} \cdot \gamma. \label{eq:F1}
\end{align}

Similarly, we calculate the expected utility of the attacker $U_A(\theta,L,a)$ when receiving signal $L$:
\begin{align}\label{U_A(t,L,a)}
&U_A(\theta,L,a)  = \sum_{\theta \in \Theta} \sum_{a \in \mathbb{A}} \mu_A(\theta|L) \cdot \sigma_A(a|L) \cdot u_A(\theta,L,a) \nonumber \\
& = a_L \cdot \mu_A(\theta_1|L) \cdot (-\alpha - c_a) + a_L \cdot \mu_A(\theta_2|L) \cdot (g\alpha -c_a) \nonumber \\
& = a_L \cdot \frac{p (1-d_1) (-\alpha - c_a) + (1-p) (1-\gamma) (g\alpha -c_a)}{p(1-d_1) + (1-p) (1-\gamma)}.
\end{align}
Maximizing $U_A(\theta,L,a)$ by $a_L$, we obtain:
\begin{subequations}\label{eq:s*}
\begin{align}
a_L^* = 1, &\,\text{when } d_1> F_2,\label{s*_a}\\
a_L^* = 0, &\, \text{when }d_1< F_2,\label{s*_b}\\
a_L^* \in  [0,1], &\,\text{when } d_1= F_2,\label{s*_c}
\end{align}
\end{subequations}
where
\begin{align}
F_2  = 1 - \frac{g\alpha -c_a}{\alpha + c_a} \cdot \frac{1-p}{p} \cdot (1-\gamma). \label{eq:F2}
\end{align}
According to the model parameter constraints (table~\ref{tab:utility_constraints}) and probability ranges, we have 
\begin{equation}\label{F1F2}
F_1>0,\, F_2<1 .   
\end{equation}
Next, consider the expected utility of the honeypot $U_D(\theta_1,m,a)$:
\begin{equation}
\begin{aligned}
&U_D(\theta_1,m,a)  = \sum_{m \in \mathbb{M}} \sum_{a \in \mathbb{A}} \sigma_D(m|\theta_1) \cdot \sigma_A(a|m) \cdot u_D(\theta_1,m,a)  \\
& = d_1 \cdot a_H \cdot (-\beta -c_d +f\alpha) + d_1 \cdot (1-a_H) \cdot (-\beta-c_d)\\
& \quad + (1-d_1)\cdot a_L \cdot (-\beta+\alpha) + (1-d_1)\cdot (1-a_L)\cdot (-\beta)\\& = d_1(a_H f\alpha-a_L \alpha-c_d)+a_L\alpha-\beta .
\end{aligned}
\end{equation}
Maximizing $U_D(\theta_1,m,a)$ by $d_1$, we obtain:
\begin{subequations}\label{eq:d1*}
\begin{align}
d_1^* = 1, &\,\text{when }a_Hf\alpha-a_L\alpha-c_d >0,\label{d1a}\\
d_1^* = 0, & \,\text{when }a_H f\alpha-a_L\alpha-c_d < 0,\label{d1b}\\
d_1^* \in [0,1], &\,\text{when }a_H f\alpha-a_L\alpha-c_d =0\label{d1c}.
\end{align}
\end{subequations}
Below, we discuss in three steps based on the different values of $a_H^*$.

\textbf{Step 1}: If $d_1<F_1$, by \eqref{r*_a}, we have $a_H^*=1$.

According to the model parameter constraints, we have 
\begin{equation}
a_H f\alpha-a_L\alpha-c_d\geq f\alpha-\alpha-c_d > 0.
\end{equation}
Then by \eqref{d1a}, $d_1^* = 1$. Because $F_2<1=d_1$, we have $a_L^*=1$ by \eqref{s*_a}.
To verify it is a NE, we should show that there is no incentive for the receiver to deviate from $a_H^* = 1$. This requires $d_1 = 1 < F_1$. Thus $p < p_1$, where $p_1$ is defined in \eqref{theta12}. Thus the equilibrium strategy is 
$\left\{ \sigma_D \!=\! \begin{bmatrix}
    1 & 0 \\ \gamma & 1-\gamma
\end{bmatrix}, \sigma_A\! =\! \begin{bmatrix}
    1 & 0\\
    1 & 0
\end{bmatrix}\right\}$. Take $d_1=1$ into \eqref{PBNE_p} and \eqref{PBNE_q}, we have $\mu_H = \frac{p}{p + (1-p) \gamma}, \mu_L = 0$. Consequently, we have the equilibrium \eqref{mix_eq1}.

\textbf{Step 2}: If $d_1>F_1$, by \eqref{r*_b}, we have $a_H^*=0$.

Because 
\begin{equation}
a_H f\alpha-a_L\alpha-c_d= -a_L\alpha-c_d < 0,
\end{equation}
and by \eqref{d1a}, we have $d_1^* = 0$. 
To verify it is a NE, we should show that there is no incentive for the receiver to deviate from $a_H^* = 0$. This requires $d_1^* = 0 > F_1$. But $F_1 > 0$. Thus it is not a NE.

\textbf{Step 3}: If $d_1=F_1$, by \eqref{r*_b}, we have $a_H^*=[0,1]$. Here we consider different cases.

(1) When $F_1 < F_2$. In order to satisfy that, $p > p_2$, where $p_2$ is defined in \eqref{theta12}.
Then $d_1 = F_1 < F_2$. By \eqref{s*_b}, $a_L^* = 0$. Because $F_1 > 0$ and $F_2< 1$ by \eqref{F1F2}, we have $0 < d_1 < 1$. Thus $a_H f\alpha- a_L\alpha - c_d = 0$.
Take in $a_L^* = 0$, then $a_H^* = \frac{c_d}{f\alpha}$.
Consequently, the equilibrium strategy is 
$\left\{ \sigma_D \!=\! \begin{bmatrix}
    F_1 & 1-F_1 \\ \gamma & 1-\gamma
\end{bmatrix}, \sigma_A\! =\! \begin{bmatrix}
    \frac{c_d}{f\alpha} & 1-\frac{c_d}{f\alpha}\\
    0 & 1
\end{bmatrix}\right\}$. Take $d_1=F_1$ into \eqref{PBNE_p} and \eqref{PBNE_q}, we have $\mu_H\! = \!\frac{p F_1}{p F_1+ (1-p) \gamma}, \mu_L \!=\!\frac{p (1-F_1)}{p (1-F_1)+ (1-p) (1-\gamma)}$. Thus, we have the equilibrium \eqref{mix_eq3}.

(2) When $1 >F_1 > F_2$. In order to satisfy that, $p_1 < p < p_2$, where $p_1$ and $p_2$ are defined in \eqref{theta12}.
Because $d_1^* = F_1 > F_2$, by \eqref{s*_a}, ~$a_L^* = 1$.
Because $0 < d_1^* < 1$, $a_H f \alpha - a_L \alpha - c_d = 0$. Take in $a_L^* = 1$, we have
$a_H^* = \frac{\alpha + c_d}{f \alpha}$. Then the equilibrium strategy is 
$\left\{ \sigma_D \!=\! \begin{bmatrix}
    F_1 & 1-F_1 \\ \gamma & 1-\gamma
\end{bmatrix}, \sigma_A\! =\! \begin{bmatrix}
    \frac{\alpha+c_d}{f\alpha} & 1-\frac{\alpha+c_d}{f\alpha}\\
    1 & 0
\end{bmatrix}\right\}$. Take $d_1=F_1$ into \eqref{PBNE_p} and \eqref{PBNE_q}, we have $\mu_H\! = \!\frac{p F_1}{p F_1+ (1-p) \gamma}, \mu_L \!=\!\frac{p (1-F_1)}{p (1-F_1)+ (1-p) (1-\gamma)}$. Thus, we have the equilibrium \eqref{mix_eq2}.

(3) When $1 = F_1 > F_2$. In order to satisfy that, $p = p_1$. Because $d_1 > F_2$, by \eqref{s*_a}, we have $a_L^* = 1$.
Because $d_1^* = F_1 = 1$, $a_H f \alpha - a_L \alpha - c_ d \geq 0$. Take in $a_L^* = 1$, we have
$a_H^* \geq \frac{\alpha + c_d}{f \alpha}$. Then the equilibrium strategy is $\left\{ \sigma_D \!=\! \begin{bmatrix}
    1 & 0 \\ \gamma & 1-\gamma
\end{bmatrix}, \sigma_A\! =\! \begin{bmatrix}
   a_H^* & 1-a_H^*\\
    1 & 0
\end{bmatrix}\right\}$, where $a_H^* \in [\frac{\alpha + c_d}{f \alpha},1]$. Take $d_1=1$ into \eqref{PBNE_p} and \eqref{PBNE_q}, we have $\mu_H\! = \!\frac{p}{p + (1-p) \gamma},\mu_L \!=\! 0$. Consequently, we have the equilibrium \eqref{mix_eq4}.

(4) When $F_1 = F_2$. In order to satisfy that, $p = p_2$.
Because $d_1^* = F_1 = F_2$, we have $a_H^* \in [0, 1],\ a_L^* \in [0, 1]$.
Because $F_1 > 0,\ F_2 < 1$, we have $0 < d_1^* < 1$. Thus $a_H^*,\ a_L^*$ must satisfy 
$a_H^* f \alpha - a_L^* \alpha - c _d = 0$. Let $a_L^* =a_H^* f - \frac{c_ d}{\alpha}$. Since $a_L^* \in [0, 1],$ we have $a_H^*\! \in\! [\frac{c_d}{f \alpha},\! \frac{\alpha + c_d}{f \alpha}]$. Then the equilibrium strategy is $\left\{ \sigma_D \!=\! \begin{bmatrix}
    F_1& 1-F_1 \\ \gamma & 1-\gamma
\end{bmatrix}, \sigma_A\! =\! \begin{bmatrix}
   a_H^* & 1-a_H^*\\
    (a_H^* f - \frac{c_ d}{\alpha}) & (1-a_H^* f + \frac{c_ d}{\alpha})
\end{bmatrix}\right\}$. Take $d_1=F_1$ into \eqref{PBNE_p} and \eqref{PBNE_q}, we have $\mu_H\! = \!\frac{p F_1}{p F_1+ (1-p) \gamma}, \mu_L \!=\!\frac{p (1-F_1)}{p (1-F_1)+ (1-p) (1-\gamma)}$. Thus, we have the equilibrium \eqref{mix_eq5}.
\end{IEEEproof}

Theorem~\ref{thm:mixed_PBNE} establishes that, for every admissible interval of the honeypot probability \( p \), there exists a unique equilibrium, and this equilibrium can take one of three distinct forms. As \( p \) rises, the attacker systematically decreases the ratio of attack---regardless of whether the received signal is \( H \) or \( L \)---because the expected gain from attacking a normal node is no longer sufficient to offset the potential loss of striking a honeypot. From the perspective of the defender, the probability that a honeypot sends the high signal $F_1$ equals to $\frac{hg\alpha -c_a}{f\alpha + c_a} \cdot \frac{1-p}{p} \cdot \gamma$, which is proportional to the probability $\gamma$ that a normal node sends the high signal and inversely proportional to the fraction of honeypots $p$ in the network.
Moreover, the mixed‑strategy equilibria established in Theorem~\ref{thm:mixed_PBNE} subsume the pure‑strategy equilibrium identified in Theorem~\ref{thm2}.

\section{Optimal Defense Strategy Based on Mixed Strategy Equilibrium}\label{sec4}
In this section, we will discuss the optimal defense strategy based on mixed strategy equilibria given in Theorem \ref{thm:mixed_PBNE}.
For a network system, the defender cannot change the number of normal nodes $N$ and their liveness $\gamma$ but can set the number of honeypots $M$ and their action strategy $d_1$\cite{pibil2012game}. 
\begin{comment}
Therefore, the network signaling game differs from the signaling game studied in section \ref{sec3} in two aspects:
\begin{enumerate}
  \item \textbf{The goal of the defender is to maximize the network average utility.}  
  In the signaling game, the attacker arbitrarily selects one node from the entire set in every round and then decides whether to launch an attack against that node.  
  In the network game, by contrast, all nodes interact with the attacker simultaneously, and the  objective of the defender is to maximize the network average utility.  
  \item \textbf{The honeypot ratio $p$ is tunable.}  
 In the signaling game, the honeypot proportion~$p$ is exogenously fixed, whereas in the network signaling game, $p$ is a tunable variable. The value of $p$ is chosen so as to maximize the network average utility.
The defender computes, for each value of $\gamma$, the utilities corresponding to the various equilibria, selects the equilibrium that yields the highest utility, and then chooses $p$ so as to both ensure that equilibrium’s feasibility and maximize its associated utility.
\end{enumerate}
\end{comment}
%%%%%%%%%%%%%%%%%%%%%%%%%%%%%%%%%%%%%%%%%%%%%%%%%%%%%%%%%%%%%%%
For the convenience of discussion, we introduce the following network average utility $\overline{U}_{net}$.
\begin{Definition}[Network average utility]
\label{def:U_net}
For a network which includes $N$ normal nodes, when there is $M$ honeypots and the strategies of the defender and attacker are $\{\sigma_D,\sigma_A\}$, the network average utility is defined as follows:
\small
\begin{equation}\label{eq_net}
\begin{aligned}
\overline{U}_{net}& =  \frac{ M*U_D(\sigma_D,\sigma_A\mid \theta_1) +N*U_D(\sigma_D,\sigma_A\mid \theta_2)}{N} \\
 & = \frac{p}{1-p}\cdot U_D(\sigma_D,\sigma_A\mid \theta_1)+ U_D(\sigma_D,\sigma_A\mid \theta_2), 
\end{aligned}
\end{equation}
\normalsize
where $p = \frac{M}{N+M}$ represents the honeypot ratio in the network.
\end{Definition}
\begin{Remark} 
It is important to note that we use the number of normal nodes $N$ as the denominator in $\overline{U}_{net}$, rather than the total number of nodes $N+M$. This is because honeypots are an additional part of the normal node network, and the benefits and costs should be borne collectively by the normal nodes. 
Moreover, provided the network contains a sufficiently large number of nodes, the network average utility defined in \eqref{eq_net} under the mixed strategy closely approximates the true aggregate payoff of the network.
\end{Remark}

The optimal defense strategy comprises selecting the number of honeypots $M^*$ and adopting the equilibrium strategy $\sigma_D^*$ that maximizes the network average utility \eqref{eq_net}. To find this optimal defense strategy, 
it is needed to compute the network average utilities for different equilibria in Theorem \ref{thm:mixed_PBNE} and finds the equilibrium that yields the highest utility. The strategy corresponding to that equilibrium is the optimal defense strategy for the defender, as described in the following theorem. 

\begin{Theorem}[Optimal defense strategy]\label{theo_optimal}
Given the number $N$ and the strategy \(\gamma\) of the normal node, compute the following maximum problem:
\begin{equation}
j^* \in \arg\max_{j \in \{1,2,3 \}} \{\,U_{\mathrm{net},j}^*(\gamma)\},
\end{equation}
where \(U_{\mathrm{net},j}^*(\gamma)\) for \(j = 1, 2, 3\) are given as follows
\begin{equation}\label{eq1_u}
\overline{U}_{net,1}^*(\gamma) =\frac{(f\alpha-hc_a)g\alpha+(\beta+c_d)(c_a-hg\alpha)}{f\alpha+c_a}\cdot \gamma - g\alpha,
\end{equation}
\begin{equation}\label{eq2_u}
\begin{aligned}
\overline{U}_{net,2}^*(\gamma) 
  &= \Bigl[
       \bigl(\tfrac{h\,g\,\alpha - c_a}{f\,\alpha + c_a}
            - \tfrac{g\alpha - c_a}{\alpha + c_a}\bigr)
       (\alpha - \beta)
       + g\alpha
       - \tfrac{\alpha + c_d}{f}\,h\,g
     \Bigr]\gamma 
     \\ 
  &\quad{}+
     \tfrac{c_a(\beta - \alpha - g\alpha) - g\alpha\beta}
           {\alpha + c_a}\,,
\end{aligned}
\end{equation}
{\small\begin{equation}\label{eq3_u}
 \overline{U}_{net,3}^*(\gamma) =\left[\beta\left(\frac{g\alpha - c_a}{\alpha + c_a} - \frac{hg\alpha - c_a}{f\alpha + c_a}\right) - \frac{hg\,c_d}{f}\right]\gamma - \beta\frac{g\alpha - c_a}{\alpha + c_a}.
\end{equation}}
Then the defender can maximize the network average utility by setting $M^*$ honeypots and adopt equilibrium strategy $\sigma_D^*$, which is defined as 
\begin{equation}\label{eq_best_equ}
M^*=\frac{p^*_{eq,j^*}\,N}{1 - p^*_{eq,j^*}},\, \sigma_D^*=\sigma_D^{j^*},
\end{equation}
where 
$p^*_{eq,j^*} \in \{p^*_{eq,1}=p_1,\, p^*_{eq,2}=p^*_{eq,3}=p_2\}$,
\begin{equation*}
\sigma_D^{j^*} \in \{\sigma_D^{1}=\begin{bmatrix}
    1 & 0 \\ \gamma & 1-\gamma
\end{bmatrix}, \sigma_D^{2}=\sigma_D^{3}=\begin{bmatrix}
    F_1^* & 1-F_1^* \\ \gamma & 1-\gamma
\end{bmatrix}\},
\end{equation*}
$ p_1,p_2$ are defined in \eqref{theta12} and
$F_1^*= \frac{hg\alpha -c_a}{f\alpha + c_a} \cdot \frac{1-p_2}{p_2} \cdot \gamma$. 
\end{Theorem}
\begin{IEEEproof}
We need calculate the network average utilities $\overline{U}_{net}$ for different equilibria in Theorem \ref{thm:mixed_PBNE} and find the equilibrium and the optimal type probability $p^*$ which maximize $\overline{U}_{net}$.
Since equilibria \eqref{mix_eq4}--\eqref{mix_eq5} in Theorem \ref{thm:mixed_PBNE} are mixtures of equilibria in \eqref{mix_eq1}--\eqref{mix_eq3}, it suffices to calculate $\overline{U}_{net}$ of equilibria \eqref{mix_eq1}--\eqref{mix_eq3}. Define equilibria \eqref{mix_eq1}--\eqref{mix_eq3} as equilibrium (I)-(III).

\textbf{Equilibrium (I)}: When $p \in (0,p_1)$, define the equilibrium strategies as $\{ \sigma_D^1,\sigma_A^1 \}$, there is
\begin{equation}\label{eq_strategy1}
\left\{ \sigma_D^1 \!=\! \begin{bmatrix}
    1 & 0 \\ \gamma & 1-\gamma
\end{bmatrix}, \sigma_A^1\! =\! \begin{bmatrix}
    1 & 0\\
    1 & 0
\end{bmatrix}\right\}
\end{equation}
according to \eqref{mix_eq1}.
The utility of the honeypot is $U_D( \sigma_D^1,\sigma_A^1\mid \theta_1)=U_D(\theta_1, H, A)=-\beta-c_d+f\alpha$; the utility of the normal node is $U_D(\sigma_D^1,\sigma_A^1 \mid \theta_2) =\gamma\,(g\alpha - h g \alpha)\;-\;g\alpha$. According to \eqref{eq_net}, we can get 
\begin{equation}\label{eq_Unet1}
\overline{U}_{net}(\sigma_D^1,\sigma_A^1,p)=\frac{p}{1-p}\cdot U_D(\sigma_D^1,\sigma_A^1\mid \theta_1)+ U_D(\sigma_D^1,\sigma_A^1\mid \theta_2).
\end{equation}

Define $p$ which maximizes \eqref{eq_Unet1} as $p^*_{eq,1}$. Because $\frac{p}{1-p}$ is an increasing function and $U_D( \sigma_D^1,\sigma_A^1\mid \theta_1)>0$ according to Table \ref{tab:utility_constraints}, we have
\begin{equation}\label{eq1_theta}
p^*_{eq,1}=\arg\max_{p \in (0,p_1)} \overline{U}_{net}(\sigma_D^1,\sigma_A^1,p)=p_1-\delta\approx p_1,
\end{equation}
where $\delta>0$ is sufficiently small, which ensures that $p$ never reaches the critical threshold $p_{1}$ at which Equilibria~1 and~2 coexist, thereby guaranteeing that Equilibrium~1 is the unique equilibrium. 
Define the maximized value $\overline{U}_{net}(\sigma_D^1,\sigma_A^1,p^*_{eq,1})$ as $\overline{U}_{net,1}^*(\gamma)$. Take \eqref{eq1_theta} into \eqref{eq_Unet1}, then we have the value of $\overline{U}_{net,1}^*(\gamma)$ as \eqref{eq1_u}, which is a linear function of $\gamma$.

\textbf{Equilibrium (II)}: When $p \in (p_1,p_2)$, define the equilibrium strategies as $\{ \sigma_D^2,\sigma_A^2 \}$, there is
\begin{equation}\label{eq_strategy2}
\left\{ \sigma_D^2 \!=\! \begin{bmatrix}
    F_1 & 1-F_1 \\ \gamma & 1-\gamma
\end{bmatrix}, \sigma_A^2\! =\! \begin{bmatrix}
    \frac{\alpha+c_d}{f\alpha}  & 1-\frac{\alpha+c_d}{f\alpha} \\
    1 & 0
\end{bmatrix}\right\}
\end{equation}
according to \eqref{mix_eq2}, where 
$F_1= \frac{hg\alpha -c_a}{f\alpha + c_a} \cdot \frac{1-p}{p} \cdot \gamma$.
The utility of the honeypot can be computed as $U_D( \sigma_D^2,\sigma_A^2\mid \theta_1) = \alpha - \beta$; the utility of the normal node is $U_D(\sigma_D^2,\sigma_A^2 \mid \theta_2) =\gamma\Bigl(g\,\alpha - \frac{\alpha + c_d}{f}\;h\,g\Bigr)\;-\;g\,\alpha$. According to \eqref{eq_net}, we can get \begin{equation}\label{eq_Unet2}
\overline{U}_{net}(\sigma_D^2,\sigma_A^2,p)=\frac{p}{1-p}\cdot U_D(\sigma_D^2,\sigma_A^2\mid \theta_1)+ U_D(\sigma_D^2,\sigma_A^2\mid \theta_2).
\end{equation}

Define $p$ which maximizes \eqref{eq_Unet2} as $p^*_{eq,2}$. Because $\frac{p}{1-p}$ is an increasing function and $U_D( \sigma_D^2,\sigma_A^2\mid \theta_1)>0$ according to Table \ref{tab:utility_constraints}, we have
\begin{equation}\label{eq2_theta}
p^*_{eq,2}=\arg\max_{p \in (p_1,p_2)} \overline{U}_{net}(\sigma_D^2,\sigma_A^2,p)=p_2-\delta\approx p_2,
\end{equation}
where $\delta>0$ is sufficiently small. Similar to $\delta$ in \eqref{eq1_theta}, $\delta$ in this equation guarantees that Equilibrium~2 is the unique equilibrium. Define the maximized value $\overline{U}_{net}(\sigma_D^2,\sigma_A^2,p^*_{eq,2})$ as 
$\overline{U}_{net,2}^*(\gamma) $. Take \eqref{eq2_theta} into \eqref{eq_Unet2}, then we have the value of $\overline{U}_{net,2}^*(\gamma)$ as \eqref{eq2_u}, which is also a linear function of $\gamma$.

\textbf{Equilibrium (III)}: When $p \in (p_2,1)$, define the equilibrium strategies as $\{ \sigma_D^3,\sigma_A^3 \}$, there is
\begin{equation}\label{eq_strategy3}
\left\{ \sigma_D^3 \!=\! \begin{bmatrix}
    F_1 & 1-F_1 \\ \gamma & 1-\gamma
\end{bmatrix}, \sigma_A^3\! =\! \begin{bmatrix}
    \frac{c_d}{f\alpha}  & 1-\frac{c_d}{f\alpha} \\
    0 & 1
\end{bmatrix}\right\}
\end{equation}
according to \eqref{mix_eq3}, where 
$F_1= \frac{hg\alpha -c_a}{f\alpha + c_a} \cdot \frac{1-p}{p} \cdot \gamma$.
The utility of the honeypot can be computed as $U_D( \sigma_D^3,\sigma_A^3\mid \theta_1)= - \beta$; the utility of the normal node is $U_D(\sigma_D^3,\sigma_A^3 \mid \theta_2) =-\frac{h g c_d}{f}\,\gamma$. According to \eqref{eq_net}, we can get \begin{equation}\label{eq_Unet3}
\overline{U}_{net}(\sigma_D^3,\sigma_A^3,p)=\frac{p}{1-p}\cdot U_D(\sigma_D^3,\sigma_A^3\mid \theta_1)+ U_D(\sigma_D^3,\sigma_A^3\mid \theta_2).
\end{equation} 
Define $p$ which maximizes \eqref{eq3_theta} as $p^*_{eq,3}$. Because $\frac{p}{1-p}$ is an increasing function and $U_D( \sigma_D^3,\sigma_A^3\mid \theta_1)<0$, we have
\begin{equation}\label{eq3_theta}
p^*_{eq,3}=\arg\max_{p \in (p_2,1)} \overline{U}_{net}(\sigma_D^3,\sigma_A^3,p)=p_2+\delta\approx p_2.
\end{equation}
where $\delta>0$ is sufficiently small and guarantees that Equilibrium~3 is the unique equilibrium. Define the maximized value $\overline{U}_{net}(\sigma_D^3,\sigma_A^3,p^*_{eq,3})$ as $\overline{U}_{net,3}^*(\gamma)$. Take \eqref{eq3_theta} into \eqref{eq_Unet3}, then we have the value of $\overline{U}_{net,3}^*(\gamma)$ as \eqref{eq3_u}, which is also a linear function of $\gamma$.

To find the optimal defense strategy, we need to compare $U_{\mathrm{net},j}^*(\gamma)$ for $j \in \{1,2,3 \}$ according to \eqref{eq1_u}, \eqref{eq2_u} and \eqref{eq3_u}. All of them are linear functions of $\gamma$. Given the fixed $\gamma$, the equilibrium $j^*$ is most favorable for the defender if
\begin{equation*}
j^* \in \arg\max_{j \in \{1,2,3 \}} \{\,U_{\mathrm{net},j}^*(\gamma)\}.
\end{equation*}

The optimal defense type probability is $p^*_{eq,j^*}$, where $p^*_{eq,1}=p_1$ and $p^*_{eq,2}=p^*_{eq,3}=p_2$ according to \eqref{eq1_theta}, \eqref{eq2_theta} and \eqref{eq3_theta}. Given the number of the normal node is $N$, the number of honeypots should satisfy $\frac{M^*}{M^*+N}=p^*_{eq,j^*}$. Thus $M^*=\frac{p^*_{eq,j^*}\,N}{1 - p^*_{eq,j^*}}$.

The corresponding equilibrium strategy is $\sigma_D^{j^*}$, where $\sigma_D^{1}=\begin{bmatrix}
    1 & 0 \\ \gamma & 1-\gamma
\end{bmatrix}, \sigma_D^{2}=\sigma_D^{3}=\begin{bmatrix}
    F_1^* & 1-F_1^* \\ \gamma & 1-\gamma
\end{bmatrix}$ according to \eqref{eq_strategy1}, \eqref{eq_strategy2} and \eqref{eq_strategy3}. Because $p^*_{eq,2}=p^*_{eq,3}=p_2$, we have $F_1^*=\frac{hg\alpha -c_a}{f\alpha + c_a} \cdot \frac{1-p_2}{p_2} \cdot \gamma$.
\end{IEEEproof}
Based on Theorem \ref{theo_optimal}, the process to get 
the optimal number of honeypots $M^*$ and the optimal equilibrium strategy $\sigma_D^*$, 
which maximize the network average utility is summarized as follows:
\\(1)\;  Fix the normal--node parameters \(N\), \(\gamma\) and utility parameters
\(\alpha,\,\beta,\,c_a,\,c_d,\,f,\,g,\,h\).\\[4pt]
(2)\;  Compute the honeypot ratios
\(p_1,p_2\) from~\eqref{theta12}.\\[4pt]
(3)\;  Evaluate
\(\overline U_{\!net,j}^*(\gamma)\) for \(j=1,2,3\) using
\eqref{eq1_u}--\eqref{eq3_u}.\\[4pt]
(4)\;  Select
\(j^{*}\in\arg\max_{j\in\{1,2,3\}}\overline U_{\!net,j}^*(\gamma)\).\\[4pt]
(5)\; 
Set \(p^*_{eq,j^{*}}=p_1\) if \(j^{*}=1\); otherwise
\(p^*_{eq,j^{*}}=p_2\).  Then
\begin{equation*}
M^{*}=\dfrac{p^*_{eq,j^{*}}\,N}{1-p^*_{eq,j^{*}}},
\end{equation*}
\begin{equation*}
\sigma_D^{*}=
\begin{cases}
\begin{bmatrix}1 & 0\\ \gamma & 1-\gamma\end{bmatrix},
& j^{*}=1,\\[8pt]
\begin{bmatrix}F_1^{*} & 1-F_1^{*}\\ \gamma & 1-\gamma\end{bmatrix},
& j^{*}=2\text{ or }3,
\end{cases}
\end{equation*}
where $F_1^{*}= \dfrac{h g \alpha-c_a}{f\alpha+c_a}\,
\dfrac{1-p_2}{p_2}\,\gamma$.

%%%%%%%%%%%%%%%%%%%%%%%%%%%%%%%%%%%%%%%%%%%%%%%%%%%%%%%%%%%%%%%
%%%%%%%%%%%%%%%%%%%%%%%%%%%%%%%%%%%%%%%%%%%%%%%%%%%%%%%%%%%%%%%
\section{Fictitious play learning for the signaling game}\label{sec5}

\subsection{Fictitious Play Learning}
In this section, we analyze the signaling game in a discrete‐time fictitious-play learning framework and explore how the interplay of strategies drives the system toward equilibrium. Each player infers the strategy of its opponent from their past actions and subsequently optimizes its own strategy accordingly. 

\begin{Definition}[Fictitious Play]\cite{berger2005fictitious}
Consider a finite two–player normal–form game.  
For each time \(t\in\mathbb{T}\) and every player \(i\in\{1,2\}\),
\begin{enumerate}
    \item player \(i\) believes that his opponent \(-i\) is using a time–invariant mixed strategy $\hat{\sigma}_{-i}^t$ given by the empirical distribution of the opponent’s past actions $\{a_{-i}^0,a_{-i}^1,\cdots, a_{-i}^{t-1}\}$;
    \item player \(i\) selects a myopic best reply at time $t$ that maximizes his expected one–period payoff against the belief of the opponent strategy $\hat{\sigma}_{-i}^t$, i.e.,
          \[
              a_{i}^t\in\operatorname{BR}\bigl(\hat{\sigma}_{-i}^t \bigr).
          \]
\end{enumerate}
The sequence \(\bigl\{\hat{\sigma}_{1}^t,\hat{\sigma}_{2}^t \bigr\}_{t\ge 1}\) is called a fictitious–play path.
\end{Definition}

\begin{Definition}[Convergence of Fictitious Play]
An fictitious–play path \(\{\hat{\sigma}_{1}^t,\hat{\sigma}_{2}^t\}_{t\ge 1}\) converges to equilibrium if
\[
\operatorname{dist}\bigl(\{\hat{\sigma}_{1}^t,\hat{\sigma}_{2}^t\},\mathrm{NE}\bigr)\;\longrightarrow\;0 \quad\text{as }t\to\infty,
\]
where \(\mathrm{NE}\) is the set of Nash equilibria of the game and
\(\operatorname{dist}(.,.)\) denotes the Euclidean distance.
\end{Definition}
\begin{comment}
\begin{Definition}[Fictitious–Play Property (FPP)]
A game possesses the \emph{fictitious–play property} if \emph{every}
FP path converges to the Nash–equilibrium set, i.e.\ if all sequences of beliefs
generated by the fictitious–play procedure converge to equilibrium.
\end{Definition}
\end{comment}

Consider the case in which the honeypot probability $p$ and the normal-node strategy $\gamma$ are fixed and known to both players. In every round of the game, nature reselects the defender type according to $p$; the defender chooses a signal first, and then the attacker chooses an action. In each round, every player estimates the opponent’s strategy from the empirical frequency of the past actions, and then chooses a signal (or an action) from the best response set. Importantly, our iterated play differs from the standard reputation game. In the latter, the long‐lived player’s type remains constant across all periods (\cite{fudenberg1992maintaining, pei2022reputation}); by contrast, in our model the sender’s type is probabilistically redrawn each round. Consequently, as the number of iterations grows, the receiver can consistently infer the honeypot type’s signaling strategy.

The core components of the fictitious play include the following two parts.

\textbf{(I) Belief and strategy update of the attacker.}

Unlike the standard normal-form game, the signaling game involves asymmetric information: the attacker is unaware of the defender’s type and therefore cannot deduce the type-contingent strategy from observed play. Then we assume that the attacker is given the defender’s normal-node strategy $\gamma$.
Although in each play the type of sender is uncertain, the law of large numbers guarantees that, over large time iteration $t$, the empirical frequency of $H$ signals, defined as $P_{H}^t$, converges almost surely to $p*d_1+(1-p)*\gamma$. Thus the attacker forms the estimation for the strategy of the honeypot as
\begin{equation}\label{estimate_honey}
\hat{\sigma}_{D}^t(H|\theta_1)=
\frac{1}{p}(P_{H}^t-\gamma)+\gamma.
\end{equation}
Using $ p, \gamma$ and $\hat{\sigma}_{D}^t(H|\theta_1) $, the attacker updates the posterior belief $\mu_A(\theta\mid m),\theta \in \Theta, m \in \mathbb M$ by \eqref{mixed_belief}.
When receiving signal $m$, the attacker chooses an action from the best response set
\begin{equation}\label{BR_A}
\begin{aligned}
&\mathrm{BR}_A(m)=\arg\max_{a \in \mathbb{A}}\sum_{\theta \in \Theta}\mu_A(\theta\mid m)\,u_A(\theta,m,a)\\&=\arg\max_{a \in \mathbb{A}}\sum_{\theta \in \Theta}\frac{\hat{\sigma}_D(m\mid \theta)\,\Pr(\theta)}
      {\sum_{\tilde \theta \in\Theta}\hat{\sigma}_D(m\mid\tilde \theta)\,\Pr(\tilde \theta)}\,u_A(\theta,m,a).
\end{aligned}
\end{equation}

\textbf{(II) Strategy update of the defender.}

In every round of the game, the defender is a honeypot with probability $p$ and a normal node with probability $1-p$. 
Conditional on being a normal node, its signaling strategy is fixed: it sends $H$ with probability $\gamma$ and $L$ with probability $1-\gamma$. 
Conditional on being a honeypot, it updates strategy every round. Much simpler than the attacker,
the defender only need to form the estimations for the attacker strategies $\hat{\sigma}_A(A|H)$ and 
$\hat{\sigma}_A(A|L)$ by computing the empirical frequencies of past actions of the attacker following signals $H$ and $L$ respectively. Then the honeypot, whose type is $\theta_1$, chooses a signal from the best response set
\begin{equation}\label{BR_D}
\begin{aligned}
\mathrm{BR}_D(\theta_1)=\arg\max_{m \in \mathbb{M}}\sum_{a\in \mathbb{A}}\hat{\sigma}_A(a\mid m) u_D\bigl(\theta_1,m,a\bigr).
\end{aligned}
\end{equation}
    
The above fictitious play process is summarized in Algorithm 1.
\begin{algorithm}[h]
\caption{Fictitious play learning algorithm}
\begin{algorithmic}
    \STATE \textbf{Input:} Honeypot probability $p$, normal defender strategy $\gamma$, total iterations $T$, and defender/attacker payoff matrices.
\STATE Initialize $\mathrm{BR}_D(\theta_1)$ and the estimate $\hat{\sigma}_A(A|m), \mu_A(\theta_1|m)$ for $ m \in \{H,L\}$.
    \FOR{$t = 1$ to $T$}
        \STATE \textbf{(1) Generate defender type.}
       \STATE Draw $\theta \in\{\theta_1,\theta_2\}$ with $P(\theta=\theta_1)=p$.
        \STATE \textbf{(2) Choose a signal for the defender.}
        \IF{$\theta = \theta_1$} 
            \STATE pick any $m^* \in \mathrm{BR}_D(\theta_1)$ defined in \eqref{BR_D}
        \ELSE 
            \STATE choose $H$ with probability $\gamma$
            \STATE choose $L$ with probability $1-\gamma$
        \ENDIF
       \STATE \textbf{(3) Choose an action for the attacker.}
        \STATE Compute empirical frequency of $H$ signals $P_H$.
        \STATE Update estimate $\hat{\sigma}_{D}(H|\theta_1)$ by \eqref{estimate_honey} and posterior belief $\mu_A(\theta\mid m)$ by \eqref{mixed_belief}.
        \STATE Choose any $a^*\in \mathrm{BR}_A(m)$ defined in \eqref{BR_A}.
        \STATE \textbf{(4) Update $\mathrm{BR}_D(\theta_1)$ for type $\theta_1$ defender.}
      \STATE Update the estimate $\hat{\sigma}_A(A|H)$ and 
$\hat{\sigma}_A(A|L)$ by computing the empirical frequencies.
        \STATE Update the best response set $\mathrm{BR}_D(\theta_1)$ by \eqref{BR_D}.
    \ENDFOR
    \STATE \textbf{Output:} $\hat{\sigma}_D(H|\theta_1), \hat{\sigma}_A(A|m)$ for $m \in \{H,L\},$ $ \mu_A(\theta|m)$ for $ m \in \{H,L\}$ at each stage.
\end{algorithmic}
\end{algorithm}
\subsection{Convergence analysis}
In this subsection, we will analyze the convergence of the signaling game based on fictitious play learning. 
First, we convert the $\gamma$-fixed signaling game into its corresponding normal-form representation and then show that this induced normal-form game converges under the fictitious-play dynamics.

The following lemma guarantees the equivalence between the original signaling game and its induced normal-form version.
\begin{Lemma}\label{lemma_trans_game}
The normal-form representation of the signaling game $\mathcal{G}^0$ is specified by:
\begin{enumerate}[label=(\roman*)]
  \item Players: defender $D$ and attacker $A$;
  \item Pure strategy sets:
        \[
          S_D = \{\,\sigma_D:\Theta\to\mathbb{M}\,\},\qquad
          S_A = \{\,\sigma_A:\mathbb{M}\to\mathbb{A}\,\};
        \]
  \item Payoff functions: for every $(\sigma_D,\sigma_A)\in S_D\times S_A$,
  \end{enumerate}
\begin{equation}\label{matrixU_D}
    EU_D(\sigma_D,\sigma_A)=\sum_{\theta \in\Theta} \Pr(\theta)\,
u_D\!\bigl(\theta,\sigma_D(\theta),\sigma_A(\sigma_D(\theta))\bigr),
\end{equation}
\begin{equation}\label{matrixU_A}
    EU_A(\sigma_D,\sigma_A)=\sum_{\theta \in\Theta} \Pr(\theta)\,
    u_A\!\bigl(\theta,\sigma_D(\theta),\sigma_A(\sigma_D(\theta))\bigr).
\end{equation}
Then the Nash equilibria of this normal-form game are exactly the Perfect Bayesian equilibria of the original signaling game.
\end{Lemma}
Lemma 1 is a slight modification of \cite{leyton2008essentials}, where we extend the conclusion from Bayesian games to signaling games.
To prove the convergence of the fictitious play, we still need another lemma.
\begin{table*}[h]
\centering
\caption{The utility matrix of the defender in the normal-form game.}
\label{utility matrix for the defender}
\small
\begin{tabular}{c|c|c}
Strategies & H & L \\ \hline
$\{A,A\}$ & 
$\displaystyle p\bigl(-\beta - c_d + f\alpha\bigr)
   + (1-p)\,\gamma\,\bigl(-h g\alpha\bigr)
   + (1-p)(1-\gamma)\,\bigl(-g\alpha\bigr)$ 
& 
$\displaystyle p\bigl(-\beta + \alpha\bigr)
   + (1-p)\,\gamma\,\bigl(-h g\alpha\bigr)
   + (1-p)(1-\gamma)\,\bigl(-g\alpha\bigr)$ 
\\[1em]
$\{A,N\}$ & 
$\displaystyle p\bigl(-\beta - c_d + f\alpha\bigr)
   + (1-p)\,\gamma\,\bigl(-h g\alpha\bigr)$ 
& 
$\displaystyle -\,p\,\beta
   + (1-p)\,\gamma\,\bigl(-h g\alpha\bigr)$ 
\\[1em]
$\{N,A\}$ & 
$\displaystyle p\bigl(-\beta - c_d\bigr)
   + (1-p)(1-\gamma)\,\bigl(-g\alpha\bigr)$ 
& 
$\displaystyle p\bigl(-\beta + \alpha\bigr)
   + (1-p)(1-\gamma)\,\bigl(-g\alpha\bigr)$ 
\\[1em]
$\{N,N\}$ & 
$\displaystyle -\,p\,\bigl(\beta + c_d\bigr)$ 
& 
$\displaystyle -\,p\,\beta$
\end{tabular}
\end{table*}

\begin{table*}[h]
\centering
\caption{The utility matrix of the attacker in the normal-form game.}
\label{utility matrix for the attacker}
\small
\begin{tabular}{c|c|c}
Strategies & H & L \\ \hline
$\{A,A\}$ & 
$\displaystyle p\bigl(-f\alpha - c_a\bigr)
   + (1-p)\bigl[\gamma\,g\alpha\,(h-1) + g\alpha - c_a\bigr]$ 
& 
$\displaystyle p\bigl(-\alpha - c_a\bigr)
   + (1-p)\bigl[\gamma\,g\alpha\,(h-1) + g\alpha - c_a\bigr]$ 
\\[1em]
$\{A,N\}$ & 
$\displaystyle p\bigl(-f\alpha - c_a\bigr)
   + (1-p)\,\gamma\,\bigl(hg\alpha - c_a\bigr)$ 
& 
$\displaystyle (1-p)\,\gamma\,\bigl(hg\alpha - c_a\bigr)$ 
\\[1em]
$\{N,A\}$ & 
$\displaystyle (1-p)(1-\gamma)\,\bigl(g\alpha - c_a\bigr)$ 
& 
$\displaystyle p\bigl(-\alpha - c_a\bigr)
   + (1-p)(1-\gamma)\,\bigl(g\alpha - c_a\bigr)$ 
\\[1em]
$\{N,N\}$ & 
$\displaystyle 0$ 
& 
$\displaystyle 0$
\end{tabular}
\end{table*}

\begin{Lemma}\cite{berger2005fictitious},\cite{von2002computing}\label{lemma_converge}
Every discrete‐time fictitious-play path approaches equilibrium in every nondegenerate $2\times n$ game, where we call a bimatrix game non-degenerate if,
for every mixed strategy of either player,
the number of the opponent’s pure best responses
is no larger than the support size of that mixed strategy.
\end{Lemma}
Then we can give the following theorem.

\begin{Theorem}[Convergence of fictitious play]
Fix 
$\sigma_D(H\mid \theta_2)=\gamma$ and assume
$\gamma \;\neq\; \frac{b_1}{a_1+b_1}$.
For any optimal honeypot ratio $ p^{\!*} \in (0,1)$, there exists a sufficiently small $\delta>0$ such that, if 
\begin{equation}\label{eq_Convergence_defence}
p \;\in\; 
\{p^{\!*}-\delta,\;p^{\!*}+\delta\},
\end{equation}
the discrete-time fictitious-play
    path of the $\gamma$-fixed signaling game $\mathcal G^{0}$ converges to
    the unique equilibrium.
\end{Theorem}
\begin{comment}
\begin{Theorem}[Convergence of Fictitious Play]
Whatever the initial beliefs are for both players, when $\gamma \neq \frac{b_1}{a_1 + b_1}$, the defender can adopt the optimal defense strategy in \eqref{eq_best_equ} which ensures the discrete-time fictitious-play path of the $\gamma-$fixed signaling game $\mathcal{G}^0$ converge to equilibrium.
\end{Theorem}
\end{comment}
\begin{IEEEproof}
According to Lemma \ref{lemma_trans_game}, we first give the normal-form representation for the $\gamma-$fixed signaling game $G^0$. 
With $\sigma_D(H\mid \theta_2)=\gamma$ fixed, the strategy for the defender is $\{\,\theta_1\to \{H,L\}\,\}.$ Thus $S_D=\{H,L\}$ only for the type $\theta_1$ defender.
Because the attacker moves after observing the message,
its strategy specifies an action for each message, i.e., $S_A = \{\,(a_H,a_L) :\mathbb{M}\to\mathbb{A}\}= \{\{A,A\},\{A,N\},\{N,A\},\{N,N\}\}$, where $a_H,a_L \in \mathbb A$ represents choosing $a_H$ following $H$ signal and choosing $a_L$ following $L$ signal. Thus this is a $2 \times 4$ normal-form game.
According to \eqref{matrixU_D}-\eqref{matrixU_A}, the utility matrices for the defender and attacker can be computed as Tables \ref{utility matrix for the defender} and \ref{utility matrix for the attacker}.

Then we prove this $2\times 4$ game is nondegenerate.

We first prove that for every mixed strategy of the attacker, the number of pure best responses of the defender is no larger than the support size of that mixed strategy. 
Assume, for the sake of contradiction, that there exists a mixed strategy for the attacker that generates a larger set of pure best responses for the defender.  
Since the defender can have at most two pure best responses, it only happens when the attacker plays a pure strategy and the defender have two best responses, producing $2 > 1$ and violating
non–degeneracy. 
However, under the parameter restrictions
summarized in Table \ref{tab:utility_constraints}, such situation is impossible for any pure strategy the attacker plays; consequently the defender’s
best-response correspondence satisfies the non-degeneracy requirement.

Secondly, we prove that for every mixed strategy of the defender, the number of pure best responses of the attacker is no larger than the support size of that mixed strategy. This means in Table~\ref{utility matrix for the attacker}, for any mixed strategy chosen by the column player (defender), the number of the
row player’s (attacker) best pure responses does not exceed the support size of the column mixture (which equals \(1\) or \(2\)).  The key is to rule
out cases in which two or more rows are tied for the highest payoff.
Assuming the defender adopts a mixed strategy with $\sigma_D(H\mid \theta_1)=d_1 \in [0,1]$, the resulting payoffs can be expressed in Table \ref{tab:exp_payoff_linear}, where $a_1=h g \alpha-c_a,\,a_2=f \alpha+c_a,\,b_1=g \alpha-c_a,\,b_2= \alpha+c_a  .$
\begin{table}[h]
\centering
\caption{Expected utilities of the attacker when the defender has strategy $\sigma_D(H\mid \theta_1)=d_1$.}
\label{tab:exp_payoff_linear}
\small
\begin{tabular}{c|c}
\toprule
Strategies & Expected payoff with $\sigma_D(H\mid \theta_1)=d_1$ \\
\midrule
$\{A,A\}$ &
$\displaystyle
  \begin{aligned}
    &[-d_1\,a_{2}-(1-d_1)\,b_{2}-\gamma a_{1}-(1-\gamma)b_{1}]p\\
    &\quad+[\gamma a_{1}+(1-\gamma)b_{1}]
  \end{aligned}$ \\[1.2em]
\midrule
$\{A,N\}$ &
$\displaystyle\bigl[-d_1\,a_{2}-\gamma a_{1}\bigr]p
   +\gamma a_{1}$ \\[0.8em]
\midrule
$\{N,A\}$ &
$\displaystyle\bigl[-(1-\gamma)b_{1}-(1-d_1)\,b_{2}\bigr]p
   +(1-\gamma)b_{1}$ \\[0.8em]
\midrule
$\{N,N\}$ & $0$ \\
\bottomrule
\end{tabular}
\end{table}
Each expected payoff is a linear function of
the honeypot probability \(p\in(0,1)\). 
Since the defender can choose $p$, as long as \(p\) does not coincide exactly with the intersection of two (or more) expected utilities lines on the interval $[0,1]$ in Table \ref{tab:exp_payoff_linear}, non‐degeneracy is preserved. If the optimal honeypot ratio $ p^{\!*} \in (0,1)$ happens to occur at such an intersection, one may perturb \(p^{\!*}\) by a small amount \(\delta\) (i.e. $p \;\in\; 
\{p^{\!*}-\delta,\;p^{\!*}+\delta\}$) so as to avoid the crossing.

Moreover, we must rule out the possibility that any two of the expected-payoff lines in Table \ref{tab:exp_payoff_linear} coincide. If such coincidence occurs, it must violate the
non-degeneracy requirement. The only pair that can possibly coincide is
the \(\{A,N\}\) row and the \(\{N,A\}\) row. If they overlap, we have
\begin{equation}
\begin{aligned}
-d_1\,a_{2}-\gamma a_{1}&=-(1-\gamma)b_{1}-(1-d_1)\,b_{2},\\
\gamma a_{1}&=(1-\gamma)b_{1}.
\end{aligned}
\end{equation}
Thus we have 
\begin{equation}\label{eq_coincide}
\gamma = \frac{b_1}{a_1 + b_1}, \sigma_D(H\mid \theta_1)=d_1 = \frac{b_2}{a_2 + b_2},
\end{equation}
We can prove that the strategy in \eqref{eq_coincide} satisfy the optimal defense strategy \eqref{eq_best_equ}, which is the equilibrium strategy. Thus we should assume that $\gamma \neq \frac{b_1}{a_1 + b_1}$ to ensure there is no lines for the expected utilities of the attacker can coincide. 
With this degeneracy removed, the attacker’s
best-response correspondence also satisfies the non-degeneracy requirement.

To summary, when \(\gamma \neq \frac{b_{1}}{a_{1} + b_{1}}\), the \(2\times4\) game is non-degenerate. By Lemma~\ref{lemma_converge}, its fictitious-play path converges to the unique equilibrium. Because this \(2\times4\) game is the normal-form representation of the \(\gamma\)-fixed signaling game \(\mathcal{G}^{0}\), the fictitious-play path of \(\mathcal{G}^{0}\) converges to the same equilibrium.
\end{IEEEproof}

To conclude, in the fictitious–play learning framework, we have established that every play path ultimately converges to an equilibrium.  Hence, by Theorem \ref{theo_optimal}, once the defender chooses the optimal number of honeypots \(M^{*}\), the fictitious-play process drives both players toward the equilibrium that is most advantageous to the defender, thereby attaining the maximal network average utility.  Note that the strategy of the defender keeps adapting according to fictitious-play updates; only in the limit does it stabilize at the equilibrium strategy \(\sigma_{D}^{\ast}\).

%%%%%%%%%%%%%%%%%%%%%%%%%%%%%%%%%%%%%%%%%%%%%%%%%%%%%%%%%%%%%%%
%%%%%%%%%%%%%%%%%%%%%%%%%%%%%%%%%%%%%%%%%%%%%%%%%%%%%%%%%%%%%%%
\section{Illustrative Examples}\label{sec6}
In this section, we present a network-security example to demonstrate how to determine the optimal defense strategy. Then we apply the fictitious play learning, showing how the strategic interactions steer the system toward the specific equilibrium which is the most favorable for the defender.

The utilities parameters of both the defender and attacker are set in table \ref{exp_para}, which satisfy the constraints in Table~\ref{tab:utility_constraints}.
\begin{table}[h]
\centering
\caption{Utility parameters used in simulation}
\label{exp_para}
\begin{tabular}{cccc} 
    \toprule
    Parameter & Value & Parameter & Value\\
    \midrule
    $\alpha$ & 10 & $\beta$ & 5\\
    $c_d$ & 80 & $c_a$ & 10\\
    $g$ & 2 & $h$ & 2\\
    $f$ & 10 & &\\
    \bottomrule
\end{tabular}
\end{table}
\subsection{Compute the optimal defense strategy}
To find the optimal defense strategy, we need to compute the network average utilities $\overline{U}_{net,j}^*(\gamma)$ corresponding to different equilibria 
according to \eqref{eq1_u}--\eqref{eq3_u}.
Using parameters in Table~\ref{exp_para}, the results are as follows
\begin{equation*}
\overline{U}_{net,1}^*(\gamma) = -\frac{175\,\gamma}{11} - 20,
\end{equation*}
\begin{equation*}
\overline{U}_{net,2}^*(\gamma) = -\frac{377\,\gamma}{22} - \frac{35}{2},
\end{equation*}
\begin{equation*}
\overline{U}_{net,3}^*(\gamma) = -\frac{679\,\gamma}{22} - \frac{5}{2}.
\end{equation*}
\begin{figure}[!t]
    \centering
    \includegraphics[width=0.4\textwidth]{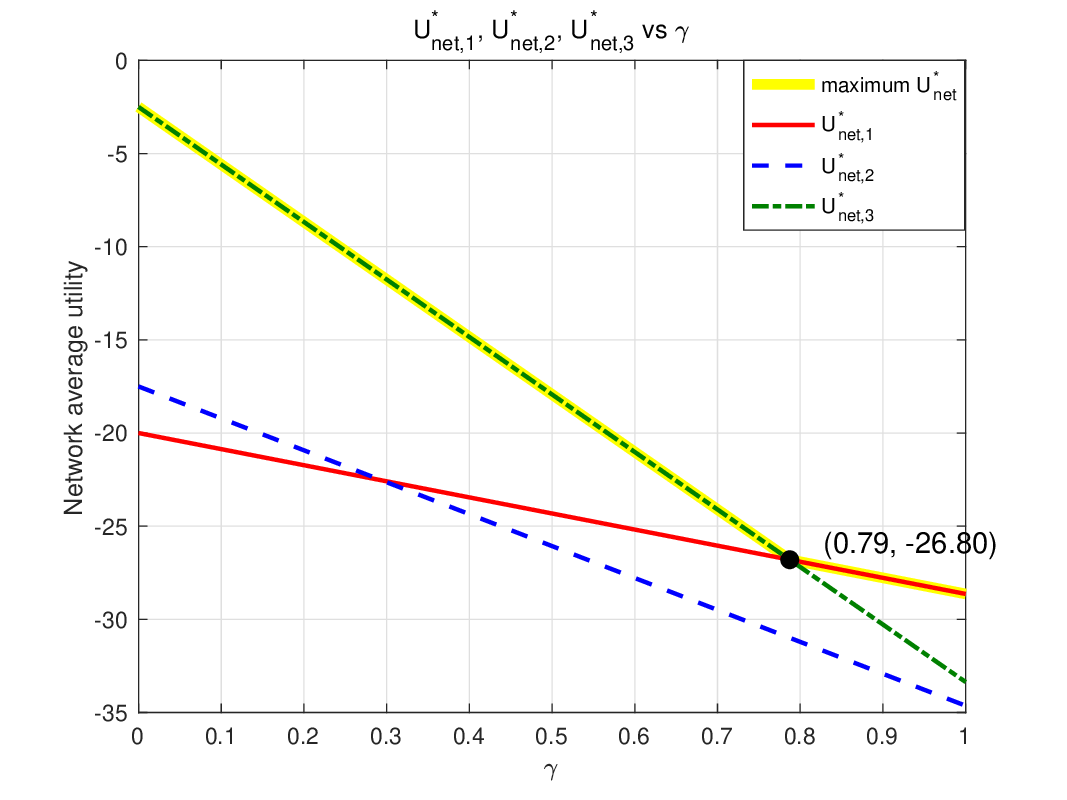}
    \caption{Optimal network average utilities for different equilibria.}
    \label{fig:evo0}
\end{figure}
Figure~\ref{fig:evo0} plots the network average utilities
\(\overline{U}_{\!net,1}^{*}(\gamma)\) (red solid),
\(\overline{U}_{\!net,2}^{*}(\gamma)\) (blue dashed), and
\(\overline{U}_{\!net,3}^{*}(\gamma)\) (green dash–dotted) as functions
of the normal–node strategy \(\gamma\).
The yellow envelope marks, for each \(\gamma\), the maximum of the three
utilities and therefore the payoff attainable by the optimal
defense strategy.
The black dot highlights the intersection point
\((\gamma,\overline{U}_{\!net}^{*})\approx(0.79,-26.8)\),
to the left of which Equilibrium 3 (green line) becomes superior and to the
right of which Equilibrium 1 (red line) becomes superior.
Thus the highest network average utility is
\begin{equation}\label{highest_net}
\max_{j \in \{1,2,3 \}} \{\,U_{\mathrm{net},j}^*(\gamma)\}=\begin{cases}
U^*_{\mathrm{net},3}(\gamma), & \gamma < 0.79,\\
U^*_{\mathrm{net},1}(\gamma), & \gamma \geq 0.79.
\end{cases}
\end{equation}

Therefore, when $\gamma < 0.79$, the defender wants to stabilize the system at equilibrium~3. According to Theorem \ref{theo_optimal}, given the number $N$ and the strategy \(\gamma\) of the normal node, the optimal defense strategy is setting $\frac{p^*_{eq,3}\,N}{1 - p^*_{eq,3}}$ honeypots and adopt equilibrium strategy $\sigma_D^3$, where
\begin{equation}\label{ex_equ3}
p^*_{eq,3}= \frac{11-5*\gamma}{33-5*\gamma}, \sigma_D^3=\begin{bmatrix}
    F_1^* & 1-F_1^* \\ \gamma & 1-\gamma
\end{bmatrix},
\end{equation}
where $F_1^* = \frac{hg\alpha -c_a}{f\alpha + c_a} \cdot \frac{1-p^*_{eq,3}}{p^*_{eq,3}} \cdot \gamma=\frac{6*\gamma}{11-5*\gamma}$.

When $\gamma \geq 0.79$, the defender prefers to stabilize the system at equilibrium~1; thus, the optimal defense strategy is setting $\frac{p^*_{eq,1}\,N}{1 - p^*_{eq,1}}$ honeypots and adopt equilibrium strategy $\sigma_D^1$, where
\begin{equation}\label{ex_equ1}
p^*_{eq,1}= \frac{3*\gamma}{3*\gamma+11}, \sigma_D^1=\begin{bmatrix}
    1 & 0 \\ \gamma & 1-\gamma
\end{bmatrix}.
\end{equation}

\subsection{Fictitious play learning simulation}
After identifying the optimal defense strategy, we apply the fictitious play learning in Algorithm 1 to show both strategies of players converge to the equilibrium. We consider two cases where $\gamma < 0.79$ and $\gamma \geq 0.79$ respectively.
\begin{figure*}[!t]           % !t
  \centering  
  %----------- 子图 (a) -----------
  \subfloat[Empirical strategies of the attacker.]{%
    \includegraphics[width=0.33\textwidth]{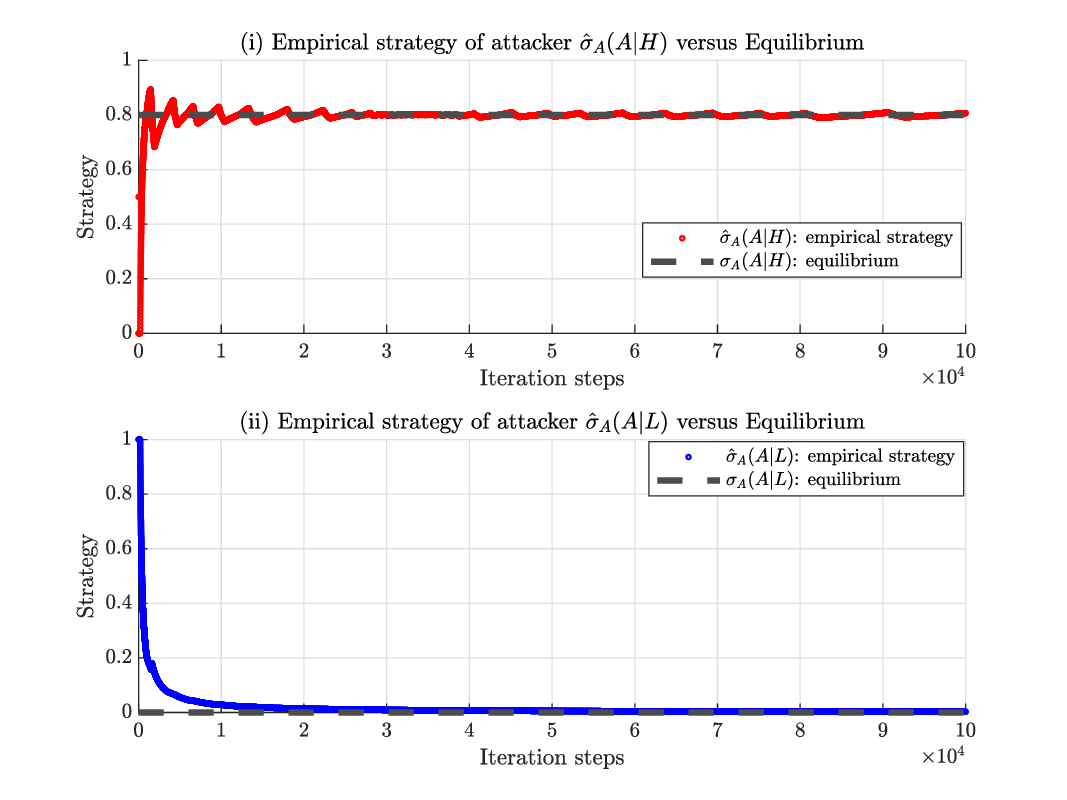}%
    \label{fig:eq3_a}}
  \hfil
  %----------- 子图 (b) -----------
  \subfloat[Empirical strategies of the defender and posteriors of the attacker.]{%
    \includegraphics[width=0.33\textwidth]{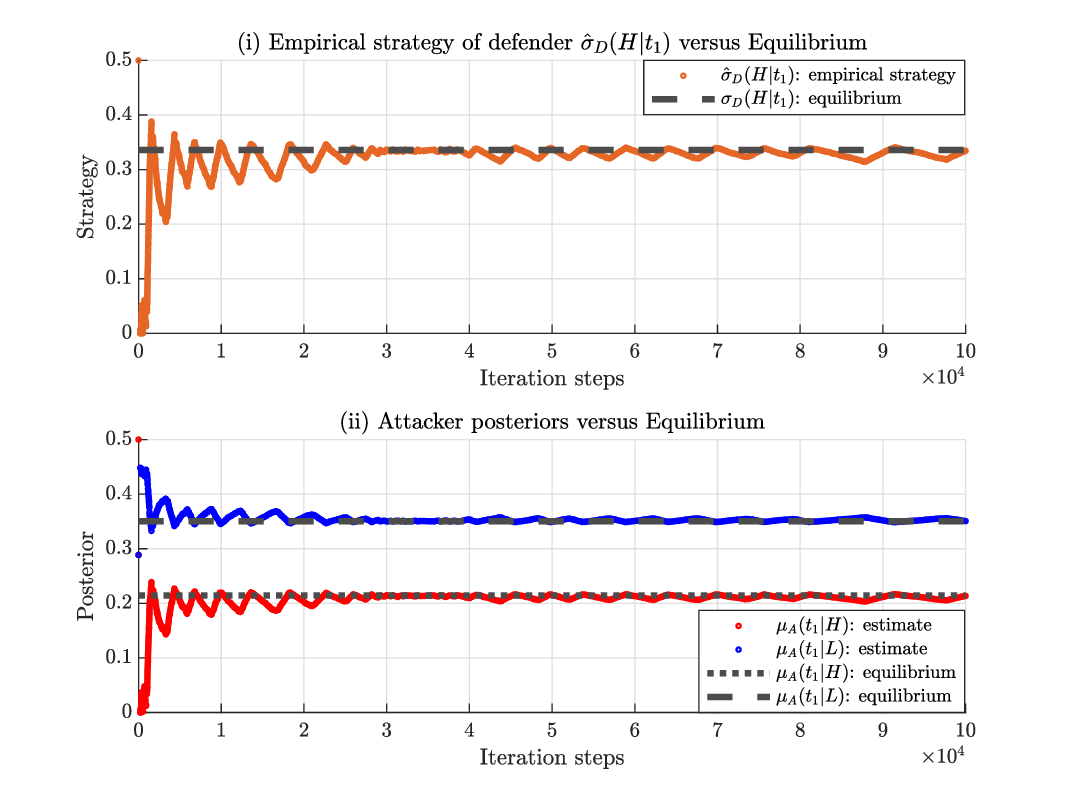}%
    \label{fig:eq3_b}}
  \hfil
  %----------- 子图 (c) -----------
  \subfloat[Evolution of network average utility $\overline{U}_{net}^*$.]{%
    \includegraphics[width=0.33\textwidth]{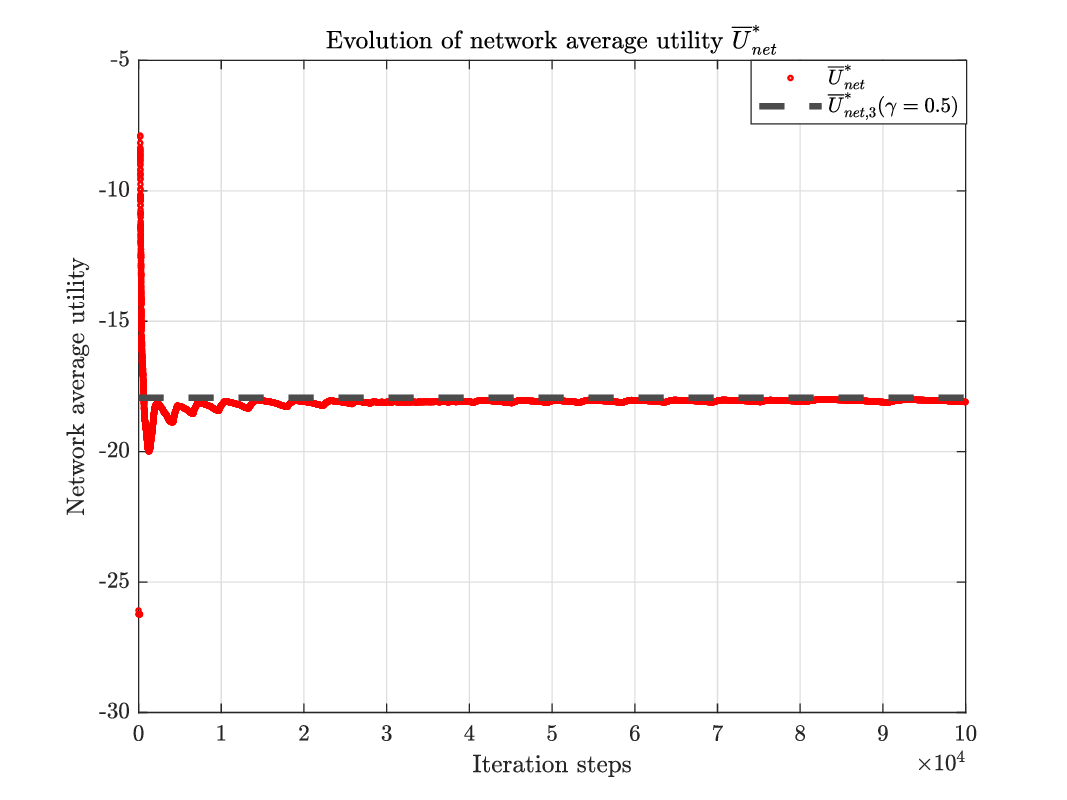}%
    \label{fig:eq3_c}}
  \caption{Beliefs and utility evolution of equilibrium (III) with $\gamma=0.5, p=p^*_{eq,3}.$}
  \label{fig:eq3}
\end{figure*}
\begin{figure*}[!t]           % !t = 尽量顶端
  \centering  
  %----------- 子图 (a) -----------
  \subfloat[Empirical strategies of the attacker.]{%
    \includegraphics[width=0.33\textwidth]{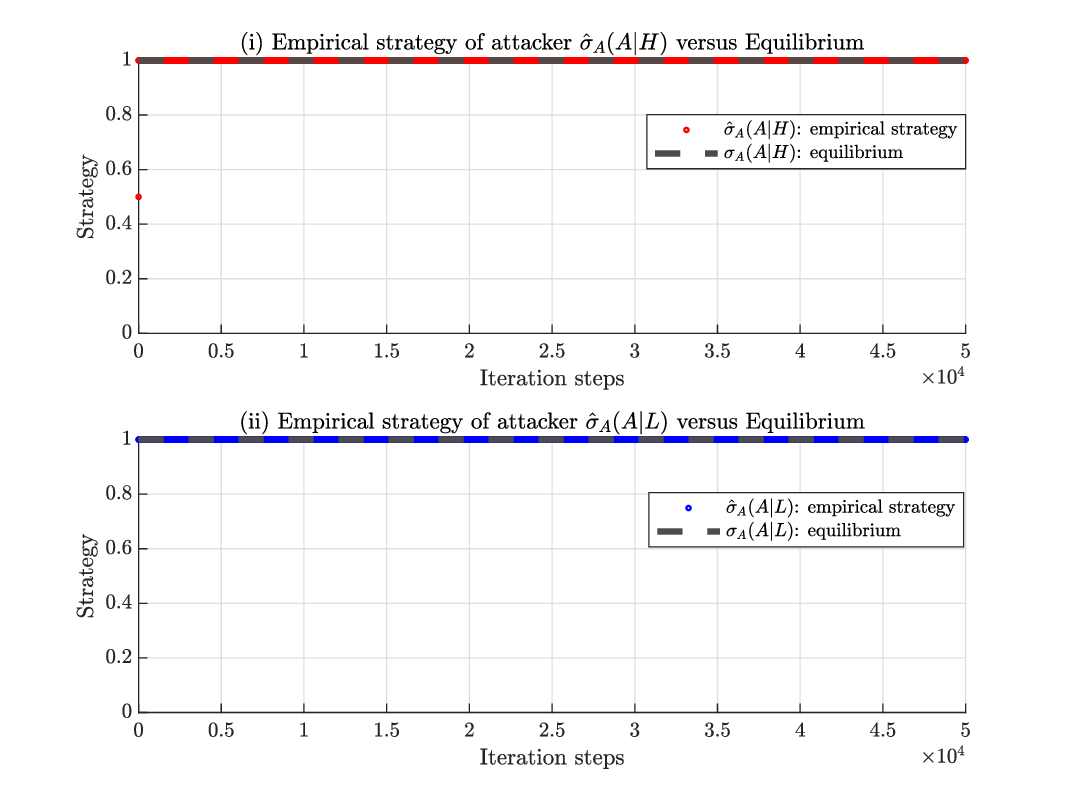}%
    \label{fig:eq1_a}}
  \hfil
  %----------- 子图 (b) -----------
  \subfloat[Empirical strategies of the defender and posteriors of the attacker.]{%
    \includegraphics[width=0.33\textwidth]{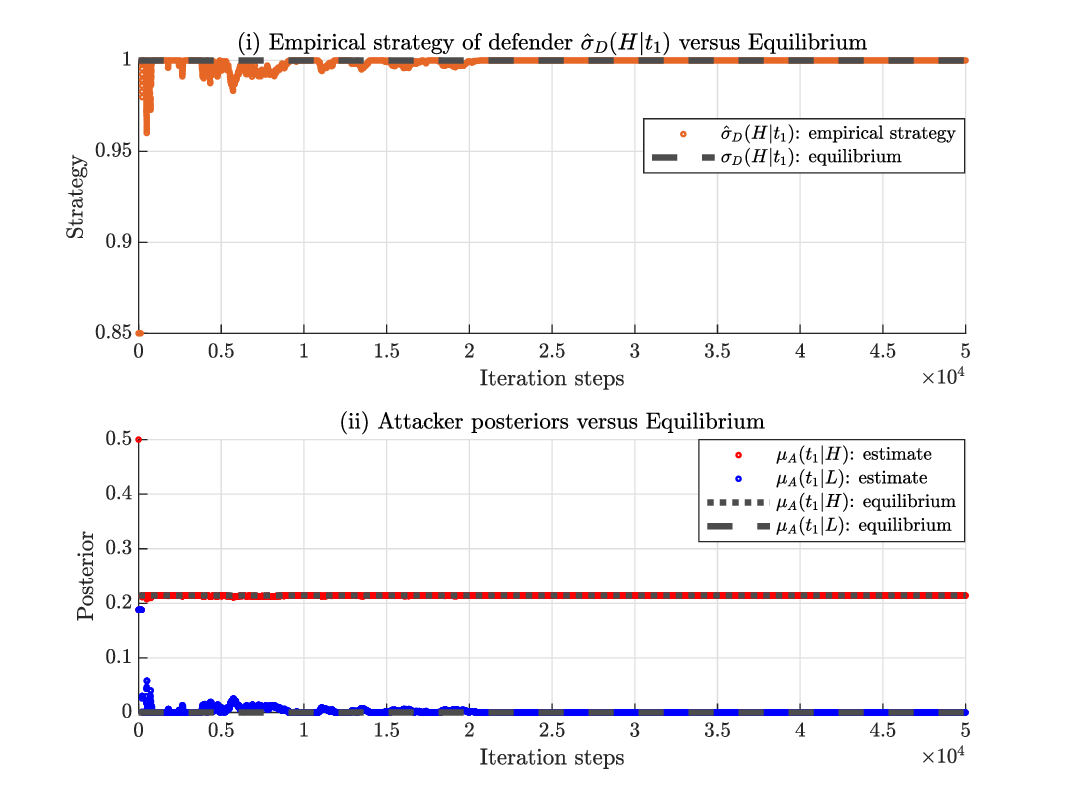}%
    \label{fig:eq1_b}}
  \hfil
  %----------- 子图 (c) -----------
  \subfloat[Evolution of network average utility $\overline{U}_{net}^*$.]{%
    \includegraphics[width=0.33\textwidth]{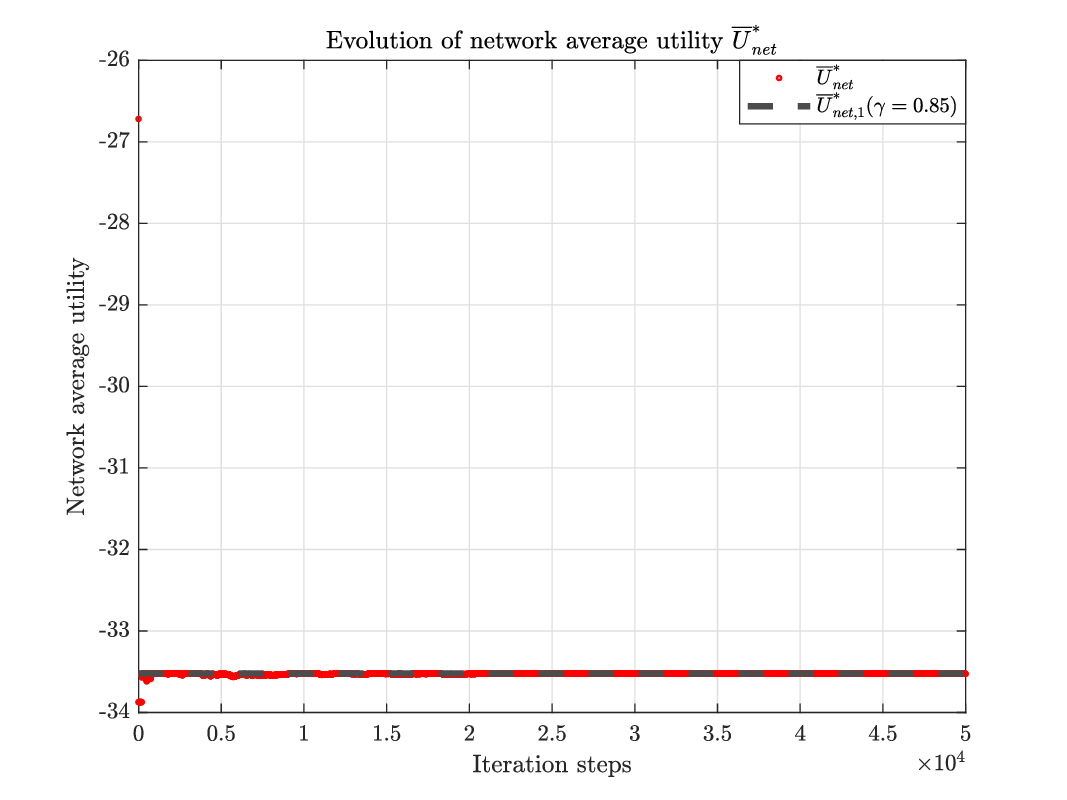}%
    \label{fig:eq1_c}}
  \caption{Beliefs and utility evolution of equilibrium (I) with $\gamma=0.85, p=p^*_{eq,1}.$}
  \label{fig:eq1}
\end{figure*}

\textbf{Case 1:} Let $\gamma = 0.5$. 

According to \eqref{ex_equ3}, the defender sets the optimal honeypot probability as \(p^*_{eq,3}+\delta \approx 0.289\), where $\delta>0$ is chosen to be sufficiently small: it
moves $p$ away from the boundary $p^*_{eq,3}$ where Equilibria~2 and~3
coincide, thereby guaranteeing complete convergence to Equilibrium~3,
yet it is tiny enough that the resulting network average utility remains virtually maximal.

The theoretical mixed–strategy equilibrium according to \eqref{mix_eq3} equals to
\begin{equation*}
\begin{aligned}
&\sigma_D^3 = 
\begin{bmatrix}
0.35 & 0.65 \\[4pt]
0.5 & 0.5
\end{bmatrix}, 
\sigma_A^3 = 
\begin{bmatrix}
0.8 & 0.2 \\[4pt]
0 & 1
\end{bmatrix}, \\[6pt]
&\mu_A(\theta_1\mid H) \approx 0.21, \quad
\mu_A(\theta_1\mid L) \approx 0.35.
\end{aligned}
\end{equation*}
And the theoretical network average utility for the equilibrium is 
$\overline{U}_{net,3}^*(\gamma=0.5) =-17.93$. 

Set total iterations $T=10^5$ and all initialized parameters equal to 0.5. Then we simulate the dynamic fictitious play learning using Algorithm 1. 
The result is shown in Figure \ref{fig:eq3}. 
Subfigure \ref{fig:eq3_a} shows that the empirical strategies of the attacker quickly stabilize at the
  equilibrium values:  
  $\hat{\sigma}_A(A\!\mid\!H)$ stabilizes at~$0.8$ (red trace) while
  $\hat{\sigma}_A(A\!\mid\!L)$ stabilizes at~$0$ (blue trace).  After a short
  transient of roughly $10^4$ iterations only small sample–noise
  fluctuations remain.

Subfigure \ref{fig:eq3_b} shows that the empirical strategy of the honeypot $\hat{\sigma}_D(H\!\mid\!\theta_1)$ (orange) quickly stabilizes at the equilibrium values $0.35$ (black dashed).  
Moreover, the posterior beliefs of the attacker ${\mu}_A(\theta_1\!\mid\!H)$ and ${\mu}_A(\theta_1\!\mid\!L)$ converge to $0.21$ (red) and $0.35$ (blue), matching the equilibrium values indicated by the dashed lines.

Subfigure \ref{fig:eq3_c} shows that the network average utility (red markers) converges to the theoretical
  value $\overline{U}^{*}_{\!net,3}(\gamma=0.5)=-17.93$ (black dashed).
  The initial overshoot is due to random start–up beliefs of the attcker and vanishes
  within $2\times10^{4}$ iterations, confirming that fictitious play
  drives the system to the defender–optimal equilibrium. The steady–state utility is marginally lower than the benchmark
because, for stability, we set
$p=p^*_{eq,3}+\delta$ rather than the exact optimum~$p^*_{eq,3}$.

\textbf{Case 2: } Let $\gamma = 0.85$. 

By \eqref{ex_equ1}, the defender uses the optimal honeypot probability \(p^*_{eq,1}-\delta \approx 0.178\), where the same
small perturbation $\delta$ moves $p$ off the boundary
$p^*_{eq,1}$ (at which Equilibria 1 and 2 coincide) and thus guarantees
exclusive convergence to Equilibrium 1.

The resulting mixed-strategy equilibrium by \eqref{mix_eq1} is
\begin{equation*}
\begin{aligned}
&\sigma_D^3 = 
\begin{bmatrix}
1 & 0 \\[4pt]
0.85 & 0.15
\end{bmatrix}, 
\sigma_A^3 = 
\begin{bmatrix}
1 & 0 \\[4pt]
1 & 0
\end{bmatrix}, \\[6pt]
&\mu_A(\theta_1\mid H) \approx 0.21, \quad
\mu_A(\theta_1\mid L) = 0,
\end{aligned}
\end{equation*}
with theoretical network average utility
$\overline{U}_{net,1}^*(\gamma=0.85) =-33.52$.

Figure~\ref{fig:eq1} confirms these predictions.  
Subfigure~\ref{fig:eq1_a} shows that the empirical
strategy of the attacker almost instantaneously converges to the pure–strategy
equilibrium, reaching  
$\hat{\sigma}_A(A\!\mid\!H)=1$ (red) and
$\hat{\sigma}_A(A\!\mid\!L)=0$ (blue) after only a few iterations.
Subfigure~\ref{fig:eq1_b} confirms that the signaling
probability of the honeypot stabilizes at \(H\) with probability~1, while the
posterior beliefs of the attacker settle at the predicted values
\((0.21,\,0)\).
Finally, Subfigure~\ref{fig:eq1_c} shows the network average utility
rapidly approaching the theoretical benchmark \(-33.52\) and then
remaining virtually unchanged; the slight gap is due to the
\(p^*_{eq,1}-\delta\) perturbation and is negligible.

To summary, Cases 1-2 show that when both players conduct the dynamic fictitious play learning, their strategy will converge to the equilibrium. This further illustrates that when the defender employs the optimal defense strategy in Theorem \ref{theo_optimal}, the corresponding optimal equilibrium is attained, thereby realizing the maximum network average utility.

\section{Conclusion}\label{sec7}
This work has presented a game-theoretic foundation for proactive deception in CPSs, focus on a $\gamma$-fixed honeypot signaling game where normal nodes cannot alter their liveness.  
By treating node liveness as the signal, we derived explicit $\gamma$-PBNEs and solved a network-level optimization problem that prescribes the honeypot ratio and signaling policy maximizing the average network utility for the defender.  
A key insight is that the optimal ratio lies on one of two analytically computable thresholds and can therefore be implemented through simple computation.  

To investigate dynamic behaviour, we embedded Bayesian updates into a discrete fictitious-play scheme and proved its convergence to the defender-optimal equilibrium whenever the honeypot ratio is chosen within a small but positive neighbourhood of the analytical optimum. Simulations corroborated the theoretical findings, demonstrating rapid convergence and allowing the defender to achieve optimal utility.

In the future, several extensions are worth pursuing. Firstly, liveness should be treated as a measurable, possibly continuous variable rather than a binary value, which would allow the defender to fine-tune deception intensity. 
Secondly, spatial realism can be introduced by conditioning payoffs on the exact placement of honeypots and on structural properties of their neighbour nodes (e.g., degree centrality, service criticality), yielding a richer description of both signaling and attack surfaces. 
Thirdly, the framework should be generalized to evolving networks in which nodes and links appear or disappear over time, so that equilibrium concepts and learning dynamics operate on a time-varying topology. 
Finally, embedding data-driven components, such as reinforcement-learning agents, within the game-theoretic model could equip defenders with adaptive strategies capable of countering more sophisticated and non-stationary attack patterns.

\section*{appendix}
%\subsection{Proof of Theorem \ref{thm1}}
\subsection{Proof of Theorem \ref{thm2}}
\begin{IEEEproof}
\textbf{Step 1}: Assume that there exists a $\gamma$‑Pure PBNE when $m^*(\theta_1)=L$, i.e., $\sigma_D = \begin{bmatrix} 0 & 1 \\ \gamma & 1-\gamma \end{bmatrix}$. According to \eqref{mixed_belief} and \eqref{mu_HL}, we have $\mu_H=\frac{p}{p+ (1-p)(1-\gamma)}, \,\mu_L=0$. The expected utilities for the attacker upon receiving signal $L$ for different actions are:
$    U_A(\theta,L,A)= \sum_{\theta} \mu_A(\theta|L) u_A(\theta, L, A)
    =  \mu_L(-\alpha - c_a) + (1 - \mu_L)(g\alpha - c_a),\,
    U_A(\theta,L,N) = \sum_{\theta} \mu_A(\theta|L) u_A(\theta, L, N)= 0$.
To compare the above payoffs, we can define
$p_3 = \frac{(1-\gamma)(g\alpha-c_a)}{(1-\gamma)(g\alpha-c_a) +\alpha +c_a}$. According to the constraints on the parameters in Table~\ref{tab:utility_constraints}, $0<p_3<1$. We will consider the following cases (i)-(ii).

Case (i): When $p \leq p_3$, then $U_A(\theta,L,A) \geq U_A(\theta,L,N)$, making $A$ dominate $N$ for the attacker upon receiving signal $L$. Then examine the tendency of both players to deviate from their strategies. First, consider whether the defender has a tendency to deviate from signal $L$ knowing that the attacker chooses $A$ upon receiving $L$. Assuming the attacker chooses $A$ to signal $H$ , we compare the utility of the honeypot:
\begin{equation*}
u_D(\theta_1, H, A)=-\beta-c_d+f\alpha > u_D(\theta_1, L, A)=-\beta+\alpha,
\end{equation*}
Clearly, sending $H$ is more profitable for the honeypot, indicating a tendency to deviate from sending $L$. 

If we assume the attacker chooses $N$ to signal $H$, we compare the utility of the honeypot:
\begin{equation*}
u_D(\theta_1, H, N)=-\beta-c_d < u_D(\theta_1, L, A)=-\beta+\alpha.
\end{equation*}
Thus there is no tendency for the defender to deviate from signal $L$. It remains to consider the deviation tendency of the attacker from $N$ upon receiving signal $H$. Compare the expected utilities for the attacker upon
receiving signal $H$ for different actions, we have:
$U_A(\theta,H,A) 
    = \mu_H(-f\alpha-c_a)+(1-\mu_H)(hg\alpha-c_a),\,
    U_A(\theta,H,N)  = 0.$
Because $\mu_H =0$, we have $U_A(\theta,H,A) > U_A(\theta,H,N)$. Thus the attacker’s strategy would deviate from $N$ when receiving $H$, and the equilibrium does not hold when $p \leq p_3$.

Case (ii): When $p > p_3$, $U_A(\theta,L,A) < U_A(\theta,L,N)$. First, consider whether the defender has a tendency to deviate from signal $L$ knowing that the attacker chooses $N$ upon receiving $L$. Assuming the attacker’s best response to signal $H$ is $A$, we compare the utility of the honeypot:
\begin{equation*}
u_D(\theta_1, H, A)=-\beta-c_d+f\alpha > u_D(\theta_1, L, N)=-\beta,
\end{equation*}
which indicates a tendency to deviate from signal $L$. 

If we assume the attacker’s best response to signal $H$ is $N$, we compare the utility of the honeypot:
\begin{equation*}
u_D(\theta_1, H, N)=-\beta-c_d < u_D(\theta_1, L, N)=-\beta,
\end{equation*}
which indicates no tendency to deviate from signal $L$. It remains to consider whether the attacker has a tendency to deviate from $N$ upon receiving signal $H$. We have:
$    U_A(\theta,H,A) = \mu_H(-f\alpha-c_a)+(1-\mu_H)(hg\alpha-c_a),\,
    U_A(\theta,H,N) =  0$.
Because $\mu_H =0$, we have $U_A(\theta,H,A) \geq U_A(\theta,H,N)$. Thus the attacker’s strategy would deviate from $N$ when receiving signal $H$, and the equilibrium does not hold when $p >p_3$.
%To conclude, there is no $\gamma$‑Pure PBNE when $m^*(\theta_1)=L$.

\textbf{Step 2}: Assume that there exists a $\gamma$‑Pure PBNE when $m^*(\theta_1)=H$, i.e., $\sigma_D = \begin{bmatrix} 1 & 0 \\ \gamma & 1-\gamma \end{bmatrix}$. According to equation~\eqref{mixed_belief} and \eqref{mu_HL}, we have $\mu_H=\frac{p}{p+ (1-p)\gamma},\, \mu_L=0$. 
The expected utilities for the attacker upon receiving signal $H$ for different actions are:
$    U_A(\theta,H,A) 
    = \mu_H(-f\alpha - c_a) + (1 - \mu_H)(hg\alpha - c_a),\,
    U_A(\theta,H,N) =0$.
To compare the above payoffs, we need to define 
$    p_1 = \frac{\gamma(hg\alpha-c_a)}{\gamma(hg\alpha-c_a) +f\alpha +c_a}$ and 
consider the case (i) $p \leq p_1$ and case (ii) $p > p_1$ separately.

Case (i): Consider $p \leq p_1$, then $U_A(\theta,H,A) \geq U_A(\theta,H,N)$, making $A$ dominate $N$ upon receiving signal $H$. First, consider whether the honeypot has an incentive to deviate from sending $H$. We first assume that the attacker chooses $N$ when receiving signal $L$. The utilities are compared as follows:
\begin{equation*}
u_D(\theta_1, L, N)= -\beta < u_D(\theta_1, H, A)=-\beta -c_d +f\alpha,
\end{equation*}
thus the honeypot has no tendency to deviate from $H$. It remains to consider the deviation tendency of the attacker from $N$ upon receiving signal $L$. Compare the expected utilities for the attacker upon receiving signal L for different actions:
$    U_A(\theta,L,A) = \mu_L(-\alpha-c_a)+(1-\mu_L)(g\alpha-c_a),\,
    U_A(\theta,L,N) = 0$.
Because $\mu_L=0$, we have $U_A(\theta,L,A) \geq U_A(\theta,L,N)$. Thus the attacker has an incentive to deviate from $N$ when receiving signal $L$. Therefore, it is not an equilibrium. 

Next, we assume that the attacker chooses $A$ when receiving signal $L$. The utilities for the honeypot ($\theta_1$) are compared as follows:
\begin{equation*}
u_D(\theta_1, L, A)=-\beta +\alpha < u_D(\theta_1, H, A)=-\beta -c_d +f\alpha.
\end{equation*}
Thus the honeypot has no tendency to deviate from $H$. It remains to consider whether the attacker has a tendency to deviate from $A$ upon receiving signal $L$. We need to compare the utilities $U_A(\theta,L,A)$ and $U_A(\theta,L,N)$. Since $\mu_L=0$, we have $U_A(\theta,H,A) \geq U_A(\theta,H,N)$, resulting in a PBNE:
\begin{equation}\label{gamma_pure_eq}
\left\{ \sigma_D \!=\! \begin{bmatrix}
    1 & 0 \\ \gamma & 1-\gamma
\end{bmatrix}, \sigma_A\! =\! \begin{bmatrix}
    1 & 0\\
    1 & 0
\end{bmatrix}, \mu_H\! = \!\frac{p}{p + (1-p) \gamma}, \mu_L \!=\! 0 \right\}.
\end{equation}

Case (ii): $p > p_1$, then $U_A(\theta,H,A) \leq U_A(\theta,H,N)$, making $N$ dominates $A$ upon receiving signal $H$. First, consider whether the honeypot ($\theta_1$) has an incentive to deviate from $H$ knowing that the attacker chooses $N$ upon receiving $H$. Assume the attacker’s best action for signal $L$ is $A$. The utilities are compared as follows:
\begin{equation*}
u_D(\theta_1, L, A)=-\beta + \alpha > u_D(\theta_1, H, N)=-\beta -c_d,
\end{equation*}
thus the honeypot has a tendency to deviate from $H$, which is not an equilibrium. If we assume the attacker chooses $N$ when receiving signal $L$, the utilities of the defender are compared as follows:
\begin{equation*}
u_D(\theta_1, L, N)=-\beta > u_D(\theta_1, H, N)=-\beta -c_d,
\end{equation*}
Similarly, the honeypot also has a tendency to deviate from $H$. Therefore, when $p > p_1$, there is no $\gamma$‑Pure PBNE.

\end{IEEEproof}

\ifCLASSOPTIONcaptionsoff
  \newpage 
\fi

\bibliographystyle{IEEEtran}
\bibliography{reference}

\end{document}